\documentclass[landscape,usenatbib,usegraphicx]{mn2e}
\newcommand{\be}{\begin{eqnarray}}
\newcommand{\ee}{\end{eqnarray}}

\def\lsim{\,\lower2truept\hbox{${< \atop\hbox{\raise4truept\hbox{$\sim$}}}$}\,}
\def\gsim{\,\lower2truept\hbox{${> \atop\hbox{\raise4truept\hbox{$\sim$}}}$}\,}
\title[Dipole straylight contamination and alignment of low multipoles
]{The impact of dipole straylight contamination on the alignment of low multipoles of CMB anisotropies}
\author[A.Gruppuso, C.Burigana and F.Finelli]
{A.~Gruppuso \thanks{gruppuso@iasfbo.inaf.it}$^1$, 
C.~Burigana \thanks{burigana@iasfbo.inaf.it}$^1$ and
F.~Finelli \thanks{finelli@iasfbo.inaf.it}$^{1,2}$
\\
$^1$ INAF-IASF Bologna, Istituto di Astrofisica Spaziale e Fisica Cosmica di Bologna \\
Istituto Nazionale di Astrofisica, via Gobetti 101, I-40129 Bologna - Italy \\
$^2$ INAF-OAB, Osservatorio Astronomico di Bologna \\
Istituto Nazionale di Astrofisica,
via Ranzani 1, I-40127 Bologna - Italy
}

\begin{document}

\maketitle

\begin{abstract}
We estimate the impact of the Dipole Straylight Contamination (DSC) for the 
{\it Planck} satellite on the 
alignments of vectors associated to the low multipoles
of the pattern of the cosmic microwave background (CMB) anisotropies.
In particular we study how the probability distributions of eighteen estimators 
for the alignments change when DSC is taken into account.
We find that possible residual DSC should leave a non-negligible 
impact on low multipole alignments 
for effective values of
the fractional far sidelobe integrated response, $p$, 
larger than $\sim {\rm few} \times 10^{-3}$. 
The effect is strongly dependent on the intrinsic sky amplitude and
weakly dependent on the considered scanning strategy.
We find a decrease of the alignment probability between the quadrupole and the dipole 
and an increase of the alignment probability between the hexadecapole and the dipole
(larger is the intrinsic sky amplitude and lower is the contamination).
The remaining estimators do not exhibit clear signatures,
except, in some cases, considering the largest values of $p$ and the lowest sky amplitudes.
Provided that 
the real sidelobes of the {\it Planck} receivers in flight conditions 
will correspond to  
$p \lsim {\rm few} \times 10^{-3}$,
as realistically expected 
at least in the cosmological frequency channels,
and will be known with accuracies
better than $\sim {\rm few} \times 10$\% 
allowing for a suitable cleaning during
data reduction, {\it Planck} will be 
very weakly affected 
from DSC on the alignments of low multipoles. 

\end{abstract}


\raggedbottom
\setcounter{page}{1}
\section{Introduction}
\setcounter{equation}{0}
\label{intro}

The anisotropy pattern of the Cosmic Microwave Background (CMB), obtained
by Wilkinson Microwave Anisotropy Probe (WMAP\footnote{http://lambda.gsfc.nasa.gov/product/map/}), 
probes 
cosmological models with unprecedented precision \citep{spergel}.
Although WMAP data are largely consistent with the concordance 
$\Lambda$ Cold Dark Matter ($\Lambda$CDM) model, 
there are some interesting deviations from it,
in particular on the largest angular scales: 
a surprisingly low amplitude of the quadrupole
term of the angular power spectrum (APS), found by Cosmic Background Explorer (COBE) \citep{cobe,hinshaw96} and WMAP \citep{spergel},
and an unlikely (for a statistically isotropic random field) alignment
of the quadrupole and the octupole \citep{tegmark,copi,schwarz,land,vale,weeks,copi3y,abramo}. 
Moreover, both quadrupole and octupole 
align with the CMB dipole. 
Other unlikely alignments are present in the aforementioned papers and other low $\ell$ anomalies are described in \citet{eriksen}. 

It is still unknown if these anomalies come from fundamental physics  
or if they are the residual of some not removed astrophysical or
systematic effect.
This open question has attracted a lot of interest and many papers 
have been published about this subject in the last few years.

This paper represents a step forward of a previous work
(\cite{DSC}, henceforth BGF06) where the impact of the systematic 
effect induced by the CMB kinematic dipole signal
entering the main spillover
(Dipole Straylight Contamination, DSC)
on the APS
has been studied in particular for the forthcoming 
{\it Planck}\footnote{http://www.rssd.esa.int/planck} mission. 
Here we wish to estimate the main implications
of the same systematic effect on the issue of alignments 
of the low multipole vectors under experimental conditions (observational strategy, 
main properties of the far sidelobes)  
like those typically predicted for {\it Planck} 
in its cosmological frequency channels (at the lowest and 
highest {\it Planck} frequency channels 
Galactic straylight dominates over dipole straylight).

The measurement of the low $\ell$ pattern is affected by cosmic variance, 
foregrounds and systematic effects 
[for a discussion on Galactic straylight contamination
see e.g. \cite{burigana1,burigana2} and 
\cite{sandri} 
in the context
of the {\it Planck}
 Low Frequency Instrument (LFI) \citep{mandolesi},
\cite{lamarre} in the context
of the {\it Planck} High Frequency Instrument (HFI) \citep{hfi},
and \cite{barnes}, \cite{naselsky1}, in the context of WMAP;
see e.g. \cite{naselsky2, naselsky3} for an analysis of 
large angular scale foreground contamination in WMAP data].

Through a simple analytical model 
[top-hat approximation for the main spillover response
\citep{gruppuso}]
and several numerical simulations 
we tackled the systematic effect induced on the APS
at low and intermediate multipoles by the CMB kinematic dipole signal
entering the main spillover (BGF06).
In that study, we analytically found that in one survey
\footnote{With "number of surveys" we mean "number of full sky mappings"
consecutively realized by the satellite.} 
or in an odd number of surveys
the DSC map, described by the coefficients $a_{\ell m}^{SL}$, that turn on 
for $\ell \le 4$, is given by
\be
a_{10}^{SL}=2\sqrt{4 \pi \over 3} c_1 \alpha \, ,
\ee
\be
a_{1\pm 1}^{SL} = {1\over 2}{8 \pi \over 3}\left(\pm d_1 + i d_2\right) c_{23} \alpha \, ,
\ee
for the dipole,
\be
a_{2\pm 2}^{SL}=-\left( 4 \over 3 \right)^2 \sqrt{15 \over 32 \pi} \left(d_1 \pm 2 i d_2\right) c_{23} \, ,
\ee
for the quadrupole and
\be
a_{4\pm 2}^{SL}=-{4 \over 15} \sqrt{5 \over 2 \pi} \left(d_1 \pm 2 i d_2\right) c_{23} \, , \\
a_{4\pm 4}^{SL}=-{12 \over 225} \sqrt{35 \over 2 \pi} \left(d_1 \pm 4 i d_2\right) c_{23} \, ,
\ee
for the hexadecapole, where $c_{23}=c_2 + c_3$ and
\begin{eqnarray}
& & c_1 = \sqrt{3 / 4 \pi} \, f_{SL} \Delta \, \sin (2 \Delta) \, T_{10} \, ,\\
& & c_2 = 4 \sqrt{3 / 8 \pi} \, f_{SL}\, \Delta \, ,\\
& & c_3 = 4 \sqrt{3 / 8 \pi}\, f_{SL} \, \sin (2 \Delta) /2 \, ,\\
& & d_1 = \sin \Delta \, Re \left[ T_{11}\right] \, ,\\
& & d_2 = \sin \Delta \, Im \left[ T_{11}\right] \, ,
\end{eqnarray}
with $f_{SL}=p/(4 \Delta \sin \Delta)$, $p$ being the ratio between
the power entering the main spillover and the total power entering the receiver 
(i.e. essentially the power entering the main beam) and $\Delta$ being the angular side of the 
box that in the $(\theta,\varphi)$-plane describe the main spillover region. 
Note that the octupole is unaffected. The terms $T_{10}$, $Re \left[ T_{11}\right]$ and $Im \left[ T_{11}\right]$
are the coefficients of the kinematic dipole. 
In ecliptic coordinates 
their values in thermodynamic temperature are
\begin{eqnarray} 
& & Im \left[ T_{11} \right] = 0.69823 \, {\rm mK} \, , \\  
& & T_{10} = -1.32225 \, {\rm mK} \, , \\ 
& & Re \left[ T_{11} \right] = 4.69963 \, {\rm mK} \, . 
\label{Tdipole}
\end{eqnarray}
The expressions for $a_{\ell m}^{SL}$ are obtained perturbatively to the
first order in the angle ($\alpha$) between the directions of the main spillover
and of the spin axis. 
Moreover, it has been supposed that the main spillover centre is located on the plane
defined by the spin axis and the telescope line of sight 
($\theta_{mb}$,$\varphi_{mb}$).
The relaxation of the latter assumption is described by the introduction of a phase 
$\beta$ that parametrize the displacement of the main spillover direction from $\theta_{mb}$. 
The above expressions hold in the case of a 
simple scanning strategy with the spacecraft spin axis always on the ecliptic plane,
i.e., in the case of {\it Planck}, for the so-called 
nominal scanning strategy (NSS) \citep{dupac}. 
The treatment of complex scanning strategies require numerical
simulations. 

Note the simple pattern at low $\ell$ due to DSC: even multipoles are modified at the leading 
order in $\alpha$ whereas odd multipoles do not change or are only weakly contaminated 
(i.e. linearly in $\alpha$). 
The introduction of $\beta$ does not change this scheme and $\beta$ appears only to the linear order in $\alpha$ 
in the odd multipoles (BGF06).
Therefore,  
for the dipole
the introduction of $\beta$ leads
to the replacements
\be
a_{10}^{SL} \rightarrow a_{10}^{SL} \cos \beta + {8 \over 3} \sqrt{3 \over 4 \pi} d_2 c_{23} \alpha \sin \beta \, ,\\
a_{1\pm 1}^{SL} \rightarrow a_{1\pm 1}^{SL} \cos \beta + {16 \over 3} \sqrt{3 \over 2 \pi} c_1 i \alpha \sin \beta \, .
\ee
In addition, for the octupole we have the following non vanishing coefficients
\be
a_{3 \pm 2}^{SL} = -{2\over 3} \sqrt{14 \over 15 \pi} \left( \pm 2 i d_1 + d_2\right) c_{23}
\alpha \sin \beta \, , 
\ee
\be
 a_{3 \pm 3}^{SL} = {16\over 9} \sqrt{7 \over 5 \pi} i c_{1} \alpha \sin \beta \, .
\ee

In order to investigate the implications of 
a non-proper subtraction of DSC on the low $\ell$ alignments for the 
{\it Planck} experiment,
we consider not only the above summarized analytical description 
(where the NSS is adopted) but also numerical simulations 
to explore the case of a cycloidal scanning strategy (CSS) (with slow precessions) 
that is beyond the analytical approximation.

We shall consider the case of a single survey (or of an odd number of surveys),
so providing upper limits to the contamination induced by this effect.
Except for the dipole,
the final DSC impact is in fact significantly reduced by considering an even 
number of complete surveys, although in a way dependent 
on the considered scanning strategy (we remember that also
in the case of an even number of surveys a remarkable effect survives to the averaging 
if one survey is not complete).

This article is organized as follows.
In Section \ref{multivectors} the multipole vectors expansion and its link with spherical 
harmonic expansion is reviewed along the line suggested 
by \cite{weeks}\footnote{http://www.geometrygames.org/Maxwell/} discussing also
the complementarity between the information contained in the APS and  
in the multipole vectors.
In Section \ref{analysis} we describe the simulations that we have performed and 
how the estimators for the statistical analysis are defined.
In Section \ref{results} we present how the probability distribution 
of the estimators change when DSC is taken into account. Finally, 
we discuss the obtained results and draw our conclusions in Section \ref{conclusion}.

\section{Multipole Vectors}
\label{multivectors}

The alignment of multipoles is better understood
by a new representation of CMB anisotropy maps
where the $a_{\ell m}$ (coefficients of the expansion over the basis of spherical harmonics)
are replaced by vectors \citep{copi}.
In particular, each multipole order $\ell$ is represented by $\ell$
unit vectors and one amplitude $A$
\be
 a_{\ell m} \leftrightarrow A^{(\ell)}, \hat u_1, \, \cdot \cdot \cdot \, , \hat u_{\ell}
\, .
\label{association}
\ee
Note that the number of independent objects is the same in the l.h.s and r.h.s. of equation (\ref{association}):
$2 \ell +1$ for $a_{\ell m}$ 
equals $3 \ell$ (numbers of components of the vectors) $+1$ (given by $A^{(\ell)}$) $-\ell$ (because there are $\ell$ constraints due to the normalization conditions of the vectors). 
One of the advantage of this representation is that from 
these unit vectors one can easily construct scalar quantities that 
are invariant under rotation. 
Note that is not equally easy to obtain scalar quantities directly 
from the $a_{\ell m}$ coefficients that, of course, depend on the 
coordinate system. 

Equation~(\ref{association}) can be understood 
starting from this observation \citep{weeks}:
if $f$ is a solution of the Laplace equation
\be
\nabla^2 f =0 
\, ,
\ee
where $\nabla^2=\partial_x^2+\partial_y^2+\partial_z^2$ in Cartesian coordinates,
then it is possible to build a new solution $f^{\prime} $ applying a directional derivative to $f$
\be
\nabla_{\vec u} f \equiv \vec u \cdot \nabla f = f^{\prime}
\, , \; \; \; \; \; \; \nabla^2 f^{\prime} =0
\, ,
\ee
with the gradient $\nabla = (\partial_x,\partial_y,\partial_z)$.
This happens because the two operators $\nabla^2$ and $\nabla_{\vec u}$ commute. 
\citet{maxwell} repeated this observation $\ell$ times considering the $1/r$ 
potential as starting solution.
Here $\vec r = (x,y,z)$ and 
$r =\sqrt{\vec r\cdot\ \vec r} = \sqrt{x^2 + y^2 + z^2}$.
In this way, one obtains
\begin{equation}
    f_\ell(x,y,z)
    = \nabla_{\vec u_\ell} \cdots \nabla_{\vec u_2} \nabla_{\vec u_1} \;
    \frac{1}{r} \, .
    \label{Maxwell}
\end{equation}
Observe the simple pattern
that emerges as we apply the directional derivatives one at a
time:
\begin{eqnarray}
    f_0 &=& \frac{1}{r}\nonumber\\
    f_1 &=& 
\frac{(-1)(\vec u_1 \cdot \vec r)}{r^3}\nonumber\\
    f_2 &=& 
\frac{(3\cdot 1) (\vec u_1 \cdot \vec r)(\vec u_2 \cdot \vec r) + r^2(-\vec u_1 \cdot \vec u_2)}{r^5}\nonumber\\
    f_3 &=& 
\frac{(-5\cdot 3\cdot 1) (\vec u_1 \cdot \vec r)(\vec u_2 \cdot \vec r)(\vec u_3 \cdot \vec r) + r^2(...)}{r^7} \, .\nonumber 
\end{eqnarray}
The (...) stands for a polynomial which we do not write 
explicitly, being useless for the current purposes.

Moreover, writing $f_{\ell}$ in spherical coordinates once $r$ is set to $1$, one finds
the following property
\be
\tilde \nabla^2 f_{\ell}(1,\theta,\phi) = \ell (\ell +1) f_{\ell}(1,\theta,\phi)
\, ,
\ee 
where $\tilde \nabla^2$ is the angular Laplace operator defined as
\be
\tilde \nabla^2 =
-\left[ {1\over \sin \theta}\, \partial_{\theta} \left( \, \sin \theta \, \partial_{\theta}\right)
+ {1\over \sin^2 \theta} \, \partial_\phi^2 \right]
\, \label{anglaplace}.
\ee 
In other words $f_{\ell}(1,\theta,\phi)$ is eigenfunction of the angular part of the Laplace operator
with eigenvalue given by $\ell (\ell +1)$. 
This is nothing but the definition of spherical harmonics $Y_{\ell,m}$ \citep{sakurai}.
Therefore, for every $\ell$ we can write
\be
A^{(\ell)} f_{\ell}(1,\theta,\phi) = \sum_{m=-\ell}^{\ell} a_{\ell m} Y_{\ell m}(\theta, \phi)
\, ,
\label{identification}
\ee
where the amplitude $A^{(\ell)}$ has been inserted because of normalization.
Equation~(\ref{identification}) makes evident the association represented by 
equation~(\ref{association}).
From equation~(\ref{identification}) it is possible to write down the set of equations
that has to be solved to pass from $a_{\ell m}$ to multipole vectors.
In order to see that this set is solvable we count the equations 
and the unknowns involved in this set.
From equation~(\ref{identification}) we have $2 \ell +1$ equations (one 
equation for each independent $a_{\ell m}$
\footnote{In fact we would have $4 \ell  +1$ equation because each $\ell$ different from $0$ has a real and imaginary part. But considering that $a_{\ell m}$ with $m>0$ are related to those with $m<0$) 
through $a_{\ell m}^{\star} = (-1)^m a_{\ell -m}$ we are left with $2 \ell +1$ equations.}) plus $\ell$
equations from the normality conditions of the vectors 
(i.e. $\vec u_{i} \cdot \vec u_{i} =1 $ where $i$ runs from $1$ to $\ell$).
Therefore the total number of independent equations is $3 \ell +1$. 
This is also the number of unknowns because we have $3$ unknowns for each vector plus
$1$ given by the amplitude $A^{(\ell)}$. This shows that the set is solvable.


Unfortunately, an explicit analytical solution is possible only for $\ell =1$
and for $\ell \neq 1$ numerical method are needed
\footnote{Indeed, for $\ell=2$ it is possible to obtain the multipoles vectors
computing the eigenvectors of a symmetric and traceless tensor representing the quadrupole [see \cite{landmagueijo},\cite{dennis}].}.
%
%
For $\ell =1$ we have
\begin{eqnarray}
-A^{(1)} (\vec d \cdot \vec r) &=& \sum_{m=-1}^{1} a_{1 m} Y_{1 m}(\theta, \phi)
\, , \\
\vec d \cdot \vec d &=&1 \, .
\end{eqnarray}
Considering the expression for $Y_{1 m}(\theta, \phi)$ \citep{sakurai} we have the 
following analytical solution
\begin{eqnarray}
d_x &=& \mp a_{1 1}^{(R)}/\sqrt{a_{1 0}^2/2+((a_{1 1}^{(R)})^2+(a_{1 1}^{(I)})^2)}  \, , \label{dx}\\
d_y &=& \pm a_{1 1}^{(I)}/\sqrt{a_{1 0}^2/2+((a_{1 1}^{(R)})^2+(a_{1 1}^{(I)})^2)} \, , \label{dy}\\
d_z &=& \pm a_{1 0}/\sqrt{a_{1 0}^2+2 ((a_{1 1}^{(R)})^2+(a_{1 1}^{(I)})^2)} \, , \label{dz}\\
A^{(1)} &=& \mp {1\over 2} \sqrt{3\over \pi} 
\sqrt{a_{1 0}^2+2((a_{1 1}^{(R)})^2+(a_{1 1}^{(I)})^2)} \, ,
\label{A1}
\end{eqnarray}
where $\vec d = (d_x, d_y,d_z )$ and the labels $(R)$ and $(I)$ stand for real and imaginary part.

An elegant way of solving numerically the above set of equations 
(in order to obtain the multipole vectors expansion from the set of $a_{\ell m}$) 
is presented in \citet{kats} and in \citet{weeks} where the problem of
finding $\ell$ vectors is translated into the problem of finding
the zeros of a polynomial of degree $2\ell$.
This method has been implemented in a code 
developed by \cite{weeks}
whose use is aknowledged.

We end this section with two observations:
\begin{itemize} 
\item equation~(\ref{identification}) is invariant under 
the change of sign of an even number of $\vec u_{i}$ or 
under the change of sign of an odd number of $\vec u_{i}$ and $A^{(\ell)}$
\citep{abramo,kats}.
This ``reflection symmetry'' implies that in fact multipole vectors define only directions.
Of course the same symmetry has to be satisfied by quantities defined through the multipole vectors.
Hence the estimators introduced in Section \ref{analysis}, are sensitive only to directions of multipole vectors (or in other words, they are invariant under change of the sign of the vectors).
Notice the ambiguity of sign for the dipole in equations ~(\ref{dx}-\ref{A1}) as an example of the above mentioned reflection symmetry.

\item equation (\ref{identification}) is also invariant under the
trasformation
\be
& & a_{\ell m} \rightarrow c \, a_{\ell m} \label{almandc} \, ,\\
& & A^{(\ell)} \rightarrow c \, A^{(\ell)} \label{Aandc} \, ,
\ee
where $c$ is a constant.
Therefore the $\ell$ multipole vectors associated to $a_{\ell m}$
are the same as those that are associated to $c a_{\ell m}$.
This means that for a sufficiently large number N of random extractions of
the $2 \ell +1$ independent values of $a_{\ell m}$, the corresponding N sets of $\ell$ multipole vectors do not depend on 
$C_\ell$, the variance of $a_{\ell m}$, in  the remarkable case in which the $a_{\ell m}$ follow a Gaussian distribution~\footnote{This is true for every distribution writable as $f(a_{\ell m}/\sigma)$, 
where $\sigma$ is a parameter.}.
As a  consequence, the same applies to the estimators 
for the alignments (see Section \ref{analysis}).
For instance if $a_{\ell m}$ follow a Gaussian distribution, the information
contained in the multipole vectors is ``orthogonal'' to that contained 
in the APS.
This property is broken in the presence of a systematic effect 
altering the above symmetry. 
In this case the global effect could depend on the intrinsic value 
of $C_{\ell}^{sky}$. 
In particular, if not fully removed, DSC will provide 
a spurious deviation from Gaussianity.
\end{itemize}

\begin{figure*}

\centering

\includegraphics[width=4.3cm]{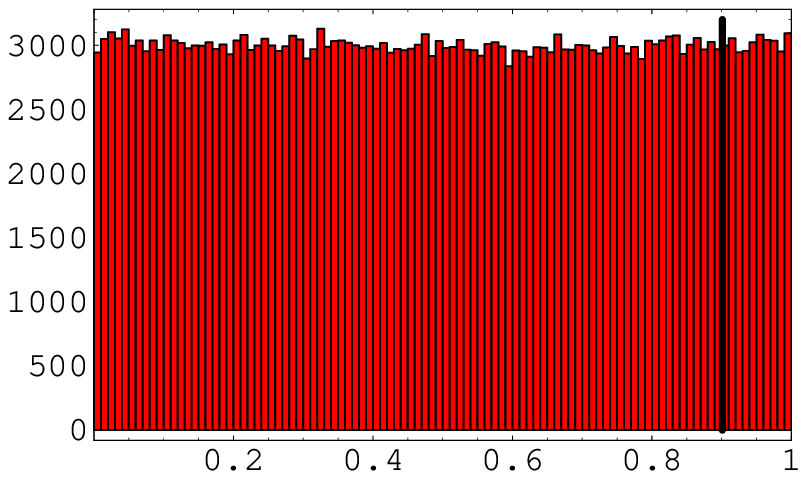}
\includegraphics[width=4.3cm]{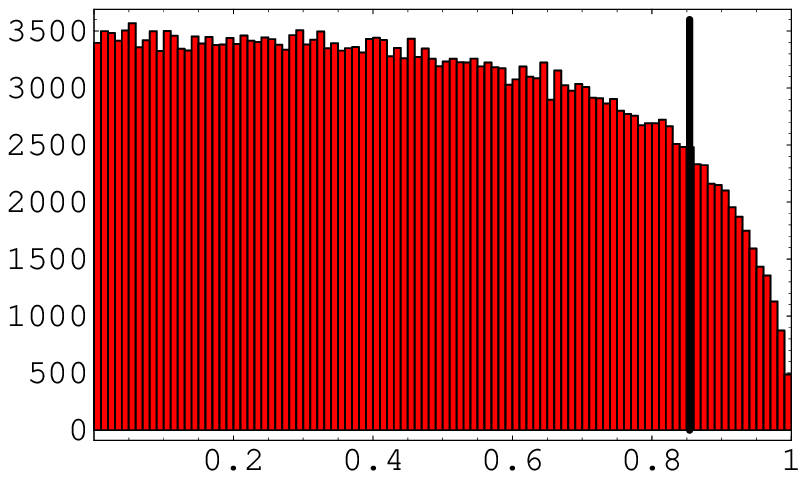}
\includegraphics[width=4.3cm]{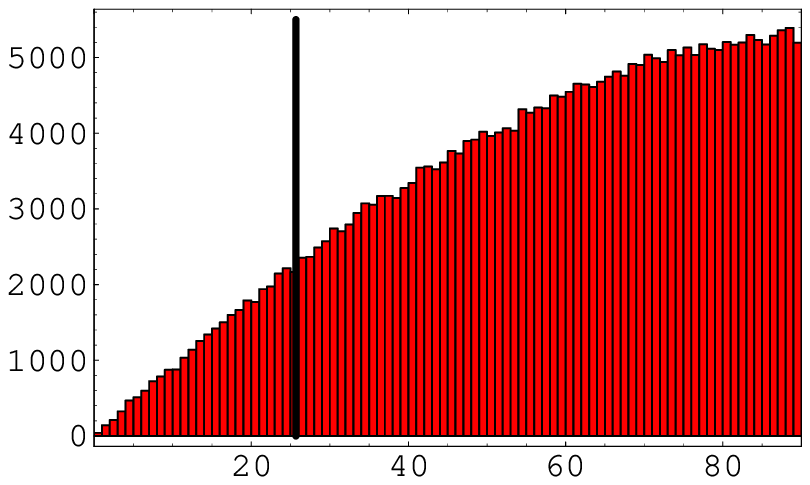}
\includegraphics[width=4.3cm]{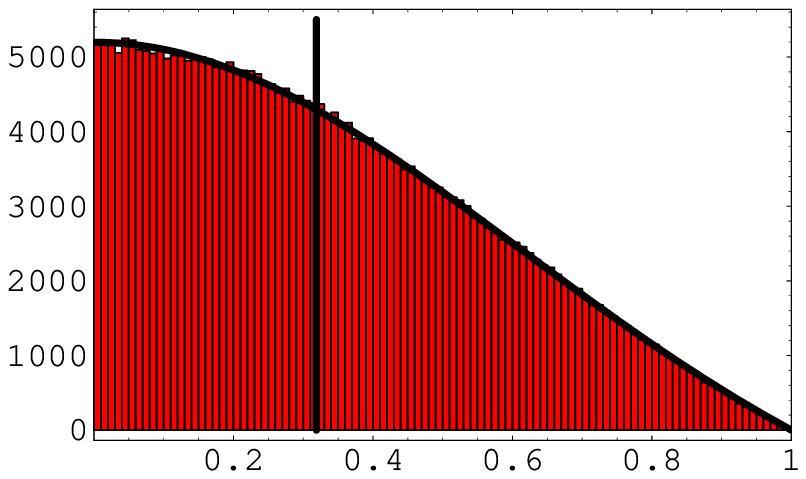}
\includegraphics[width=4.3cm]{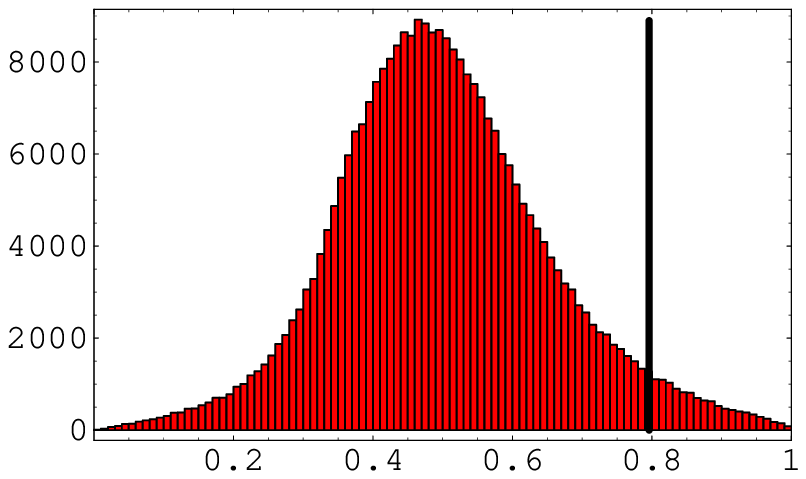}
\includegraphics[width=4.3cm]{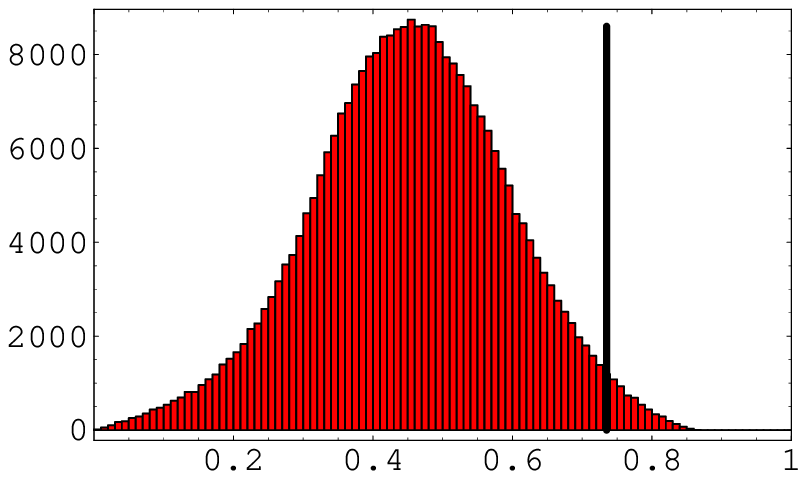}
\includegraphics[width=4.3cm]{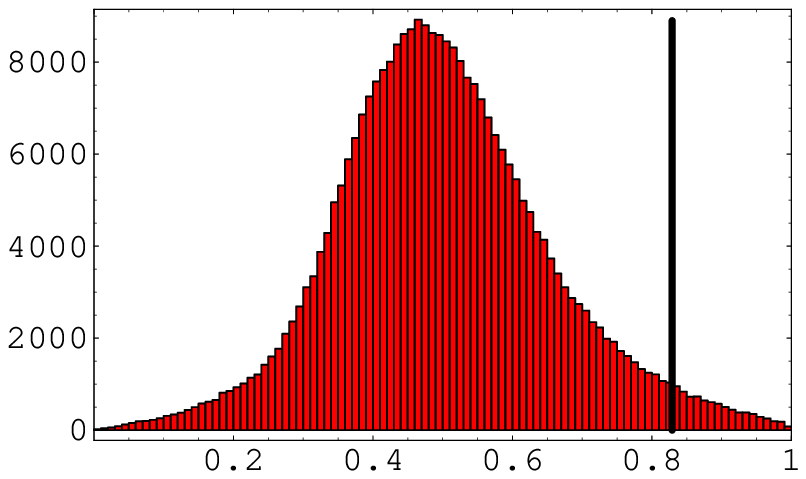}
\includegraphics[width=4.3cm]{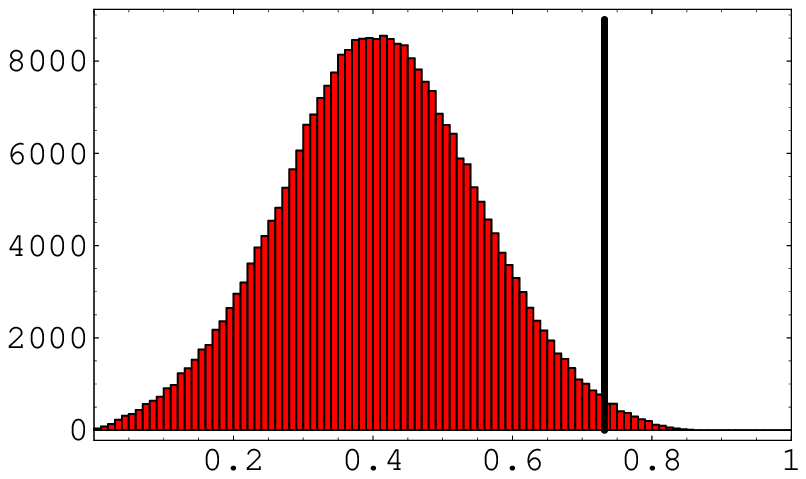}
\includegraphics[width=4.3cm]{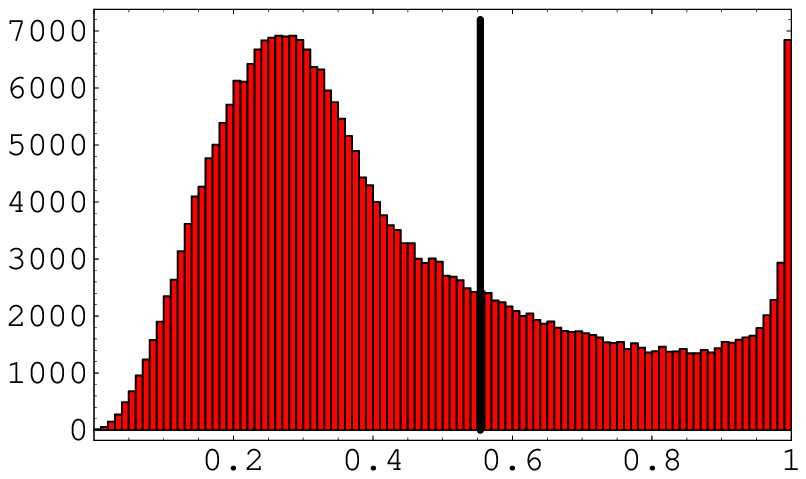}
\includegraphics[width=4.3cm]{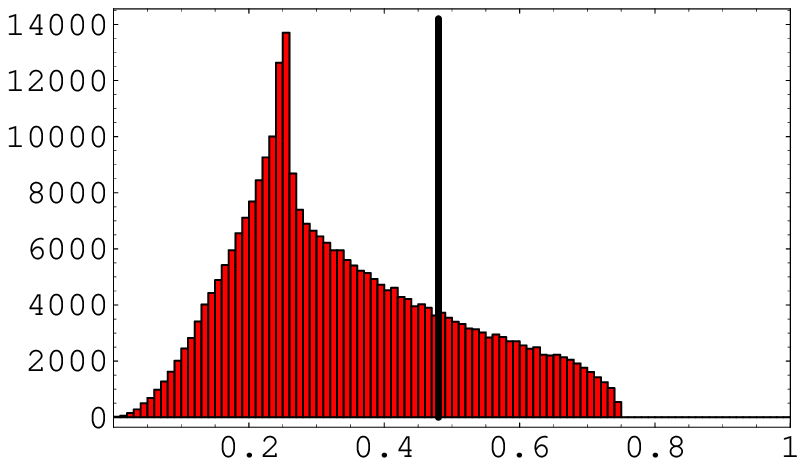}
\includegraphics[width=4.3cm]{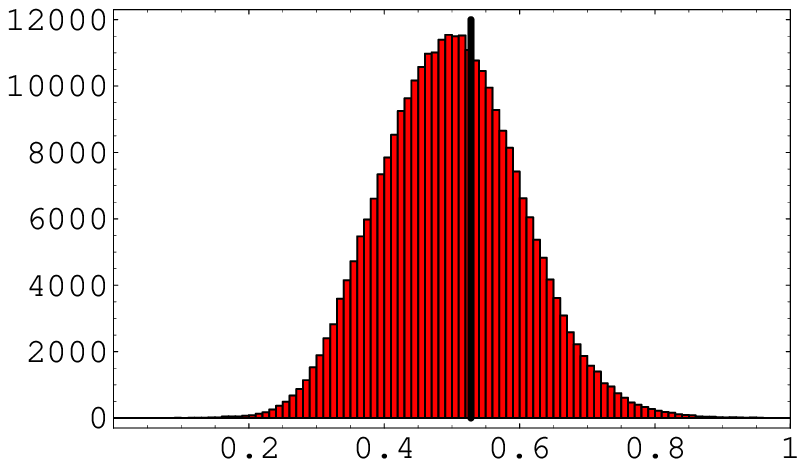}
\includegraphics[width=4.3cm]{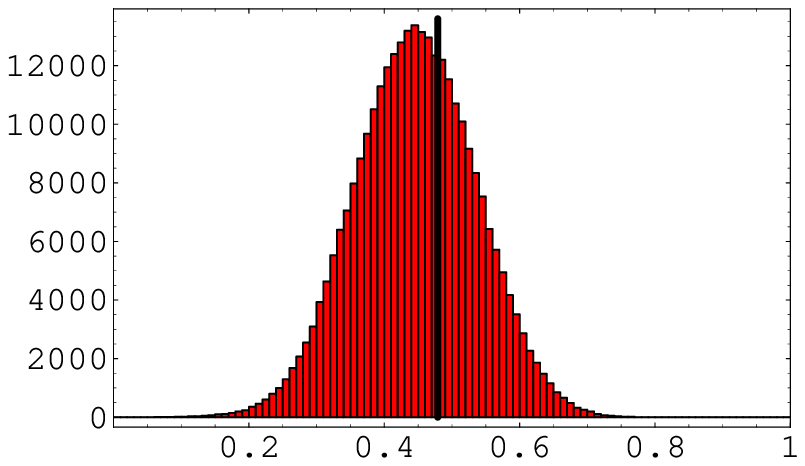}
\includegraphics[width=4.3cm]{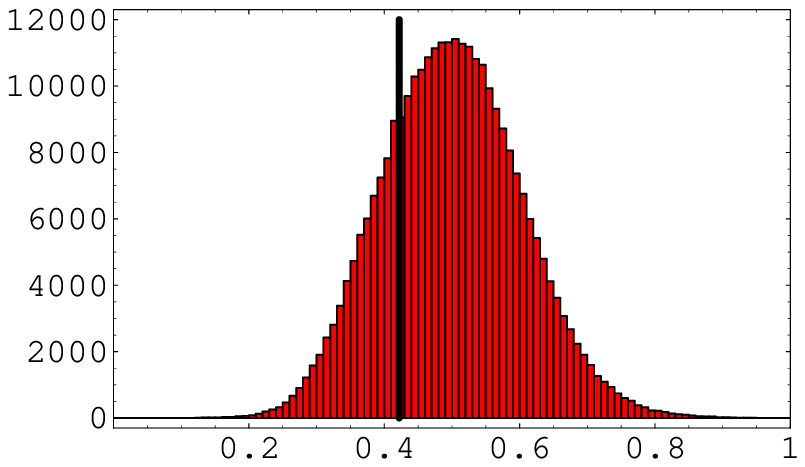}
\includegraphics[width=4.3cm]{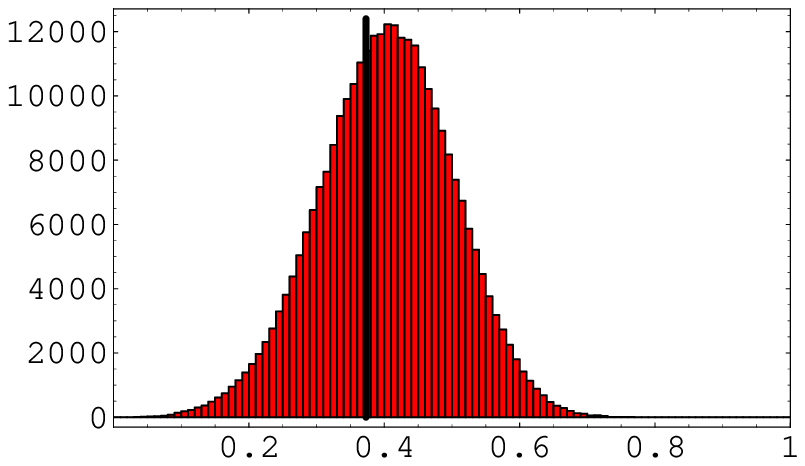}
\includegraphics[width=4.3cm]{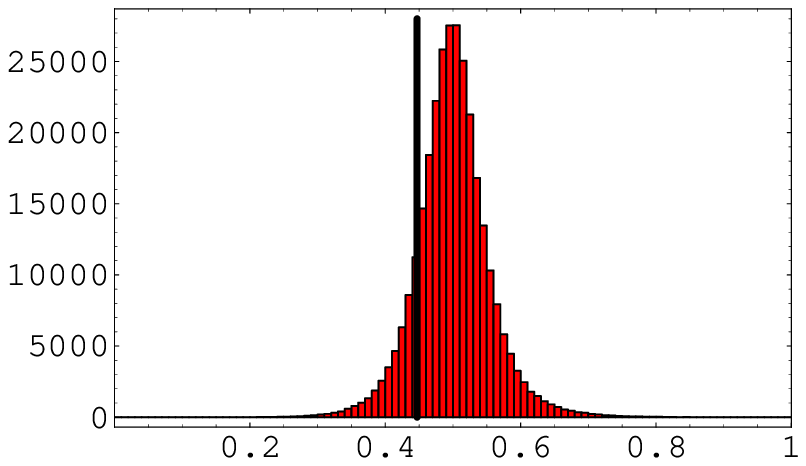}
\includegraphics[width=4.3cm]{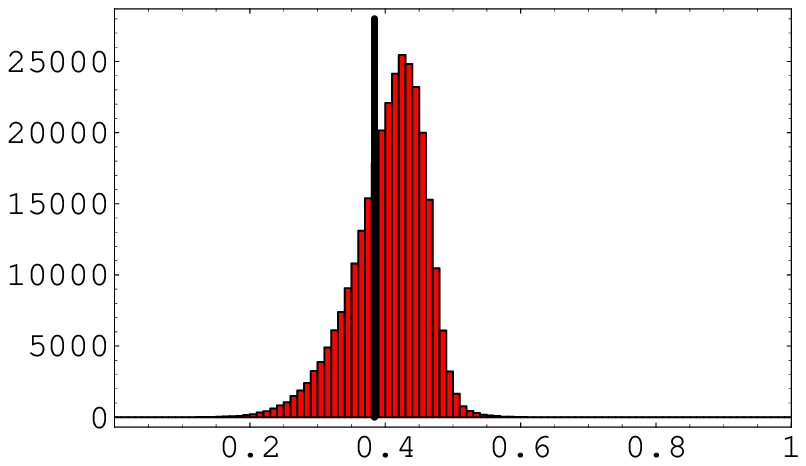}
\includegraphics[width=4.3cm]{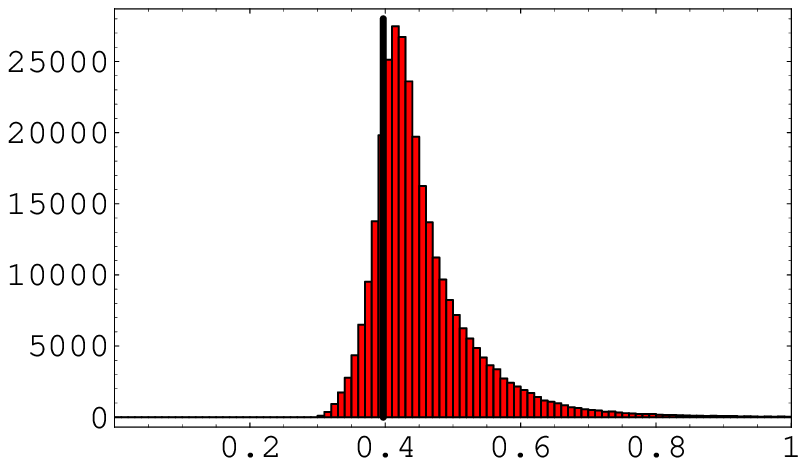}
\includegraphics[width=4.3cm]{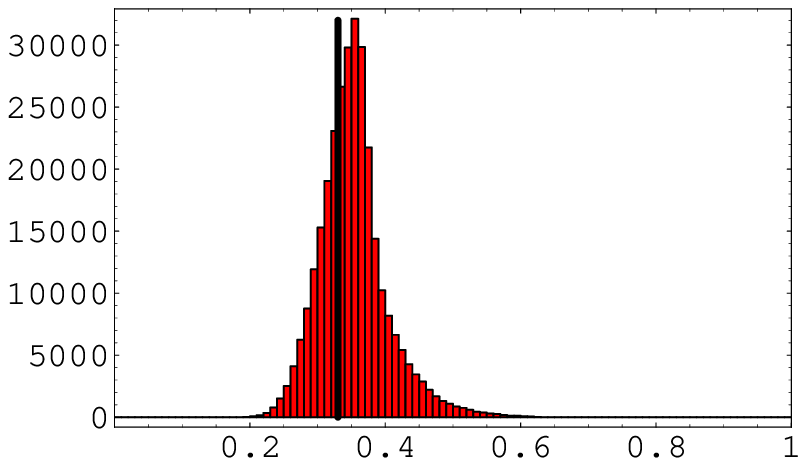}

\caption{Uncontaminated distribution for all the considered estimators.
In order from left to right and above to below: $\hat W2 $, $ W2 $, $ W2 ^{\circ}$,
$R22$, $\hat W3$, $W3$, $D23$, $S23$, $D33$, $S33$, $\hat W4$, $W4$, $D42$, $S42$, $D43$, $S43$, $D44$, $S44$. All the panels present the counts ($y$-axis) versus the statistic ($x$-axis). 
Vertical lines stand for WMAP value reported in Table 1. For $R22$ we overplot the
analytic distribution. See also the text.}

\label{fig1}

\end{figure*}

\begin{table*}
\begin{tabular}{ccc}
\begin{tabular}{ccc}
\hline
\hline
Estimator & WMAP value & Probability \%\\
\hline
$\hat W2 $ & $0.901 $ & $9.96$ \\
\hline
$W2 ^{\circ}$ & $25.68 ^{\circ}$ & $9.94$\\
\hline
$ W2 $ & $0.854$ & $8.33$\\
\hline
$R22$ & $0.319$ & $48.04$\\
\hline
$\hat W3$ & $0.796$ & $3.92$\\
\hline
$W3$ & $0.735$  & $2.03$\\
\hline
$D23$ & $0.829$ & $2.74$\\
\hline
$S23$ & $0.732$ & $0.96$\\
\hline
$D33$ & $0.554$ & $27.49$\\
\hline
\end{tabular}

&

\begin{tabular}{ccc}
\hline
\hline
Estimator & WMAP value & Probability \% \\
\hline
$S33$ & $0.480$ & $21.76$\\
\hline
$\hat W4$ & $0.528$ & $38.73$\\
\hline 
$W4$ & $0.479$ & $36.01$\\
\hline
$D42$ & $0.422$ & $76.82$\\
\hline
$S42$ & $0.373$ & $62.80$\\
\hline
$D43$ & $0.447$ & $86.43$\\
\hline
$S43$ & $0.384$ & $69.75$\\
\hline
$D44$ & $0.397$ & $82.03$\\
\hline
$S44$ & $0.330$ & $68.70$\\
\hline
\end{tabular}

\end{tabular}
\caption{WMAP values for the considered estimators adopting the CMB component in the V band
of the WMAP 3 year release (see the text for further detailes). 
The percentages are the probability that a random map would have values for the estimators
{\it larger} than those observed in the V band by WMAP. Only for $W2 ^{\circ}$ the percentage represents the probability to extract randomly a value {\it lower} than that obtained from the above mentioned data set.}
\label{one}
\end{table*}

\section{Analysis}
\setcounter{equation}{0}
\label{analysis}

In order to study the effect of DSC on the alignment of low multipole vectors
we have extracted $3 \times 10^5$ sky realizations for two different APS amplitudes
corresponding to a concordance $\Lambda$CDM model and to WMAP.
In other words, we have extracted $a_{\ell m}^{sky}$ for $\ell =2$, $\ell=3$
and $\ell =4$ from a Gaussian distribution with zero mean and variance given by $C_2$, $C_3$, $C_4$ and 
each extraction has been contaminated by $a_{\ell m}^{SL}$.
In particular defining $\Delta T_{\ell}= \ell (\ell +1) C_{\ell}/(2 \pi)$ we 
have taken $\Delta T_2= 1250 \, \mu$K$^2$, $\Delta T_3=1150 \, \mu$K$^2$ and $\Delta T_4= 1110\, \mu$K$^2$ for the 
$\Lambda$CDM case and 
$\Delta T_2=211 \, \mu$K$^2$, $\Delta T_3=1041 \, \mu$K$^2$ and 
$\Delta T_4=731 \, \mu$K$^2$ for the WMAP-like amplitude case
\citep{hinshaw}. 
Moreover, we set $\beta=0$ 
for sake of simplicity
and considered $p=1/1000,4/1000,7/1000,10/1000$, i.e. in a range 
representative of current CMB anisotropy space experiments like WMAP 
and {\it Planck} (see e.g. \cite{barnes} and \cite{sandri}).
The other parameters are freezed to the following values
$\alpha=\pi/18$, $\Delta=\pi/10$, for sake of simplicity.
The coefficients $a_{\ell m}^{sky}$ and $a_{\ell m}^{sky}+a_{\ell m}^{SL}$ are then 
transformed to multipole vectors through the Weeks' code.

As written above, not only the analytical expressions for $a_{\ell m}^{SL}$ have been used
to describe DSC but also numerical simulations have been performed
in order to consider cases beyond the analytical approximations.
In particular, we have numerically taken into account the case of 
a CSS with slow precessions 
as described 
for example in \cite{dupac},
assuming a period $T=6$ months and a semi-amplitude of $10^\circ$,
and considering $\alpha = 10^\circ$ and $\beta=60^\circ$.
In this case, we parametrize the main spillover response according to the Gaussian 
approximation. With reference to \S~3.3 of BGF06,
we note that in the remarkable case of pencil beam
the top-hat and Gaussian approximations are 
essentially equivalent for small $\Delta$.
Differently, the two cases are not equivalent 
for general values of $\alpha$.
On the other hand,
it easily to show that in the case $\alpha = 0$ they are 
equivalent up to second order in $\sigma$ and $\Delta$ provided that 
$\sigma^2=\Delta^2/3$ (and $b=p$). We then adopt here this relation
to define the beamwidth, $\sigma$, of the main spillover in the 
Gaussian approximation given the chosen value of $\Delta$ in the 
top-hat approximation used in the analytical approach.

We present in Section \ref{results} how the distributions of the eighteen estimators 
change when the DSC is properly taken into account.
The considered estimators are: 
for the alignment quadrupole-dipole
\be
W2 &=& |\vec q \cdot \hat d | \, ,\\
\hat W2 &=& |\hat q \cdot \hat d | \, ,\\
W2 ^{\circ} &=&\arccos (|\hat q \cdot \hat d|) 180/\pi
\, ,
\ee
for the self alignment of the quadrupole
\be
R22 = |\hat q_{21} \cdot \hat q_{22} | \, ,
\ee
for the alignment octupole-dipole
\be
W3 &=& \sum_{i=1}^3|\vec o_i \cdot \hat d | /3 \, ,\\
\hat W3 &=& \sum_{i=1}^3|\hat o_i \cdot \hat d | /3
\, ,
\ee
for the alignment quadrupole-octupole
\be
S23 &=& \sum_{i=1}^3|\vec q \cdot \vec o_i| /3  \, ,\\
D23 &=& \sum_{i=1}^3|\hat q \cdot \hat o_i| /3
\, ,
\ee
for the self-alignment of the octupole
\be
S33 &=& \sum_{i=1,j>i}^3|\vec o_i \cdot \vec o_j| /3 \, , \\
D33 &=& \sum_{i=1,j>i}^3|\hat o_i \cdot \hat o_j| /3
\, ,
\ee
for the alignment hexadecapole-dipole
\be
W4 &=& \sum_{i=1}^6 |\vec e_i \cdot \hat d| /6 \, ,\\
\hat W 4 &=& \sum_{i=1}^6 |\vec e_i \cdot \hat d| /6 \, ,
\ee
for the alignment hexadecapole-quadrupole
\be
S42 &=& \sum_{i=1}^6 |\vec e_i \cdot \vec q| /6 \, ,\\
D42 &=& \sum_{i=1}^6 |\hat e_i \cdot \hat q| /6  \, ,
\ee
for the alignment hexadecapole-octupole
\be
S43 &=& \sum_{i=1}^6 \sum_{j=1}^3 |\vec e_i \cdot \vec o_j| /18 \, ,\\
D43 &=& \sum_{i=1}^6 \sum_{j=1}^3 |\hat e_i \cdot \hat o_j| /18  \, ,
\ee
and for the self-alignment of the hexadecapole
\be
S44 &=& \sum_{i=1,j>i}^6 |\vec e_i \cdot \vec e_j| /15 \, ,\\
D44 &=& \sum_{i=1,j>i}^6 |\hat e_i \cdot \hat e_j| /15  \, ,
\ee
where the symbol ``hat'' stands for a vector with norm equal to $1$ and where
the ``area vectors'' are defined as
\be
\vec q = \hat q_{2 1} \times \hat q_{2 2} \, , \\
\vec o_1 = \hat o_{3 2} \times \hat o_{3 3} \, , \\
\vec o_2 = \hat o_{3 3} \times \hat o_{3 1} \, ,\\
\vec o_3 = \hat o_{3 1} \times \hat o_{3 2} \, ,\\
\vec e_1 = \hat e_{4 1} \times \hat e_{4 2} \, , \\
\vec e_2 = \hat e_{4 1} \times \hat e_{4 3} \, ,\\
\vec e_3 = \hat e_{4 1} \times \hat e_{4 4} \, , \\
\vec e_4 = \hat e_{4 2} \times \hat e_{4 3} \, , \\
\vec e_5 = \hat e_{4 2} \times \hat e_{4 4} \, ,\\
\vec e_6 = \hat e_{4 3} \times \hat e_{4 4} \, ,
\ee
with $\hat q_{2 j}$ representing the two normalized multipole vectors ($j=1,2$) associated to the quadrupole, 
$\hat o_{3 j}$ representing the three normalized multipole vectors ($j=1,2,3$) associated to the octupole and $\hat e_{4 j}$ representing the four normalized multipole vectors ($j=1,2,3,4$)
associated to the hexadecapole. 
Notice that all the estimators but $W2 ^{\circ}\in [0^{\circ},90^{\circ}]$, belong to the interval $[0,1]$ and contain absolute values 
in order to make them invariant under the reflection symmetry
discussed in Section \ref{multivectors}.
Notice also that $R22$ is the unique estimator whose distribution is 
analytically known [see \cite{landmagueijo},\cite{dennis}]:
\be
p(x)= 27 {(1-x^2)\over{(x^2+3)^{5/2}}}
\, ,
\label{anforr22}
\ee
where $x=R22$.

\begin{figure*}

\centering

\includegraphics[width=3.5cm]{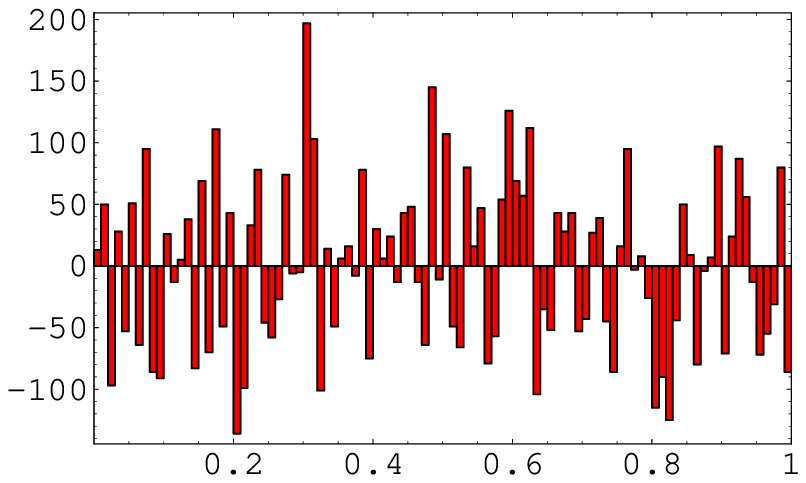}
\includegraphics[width=3.5cm]{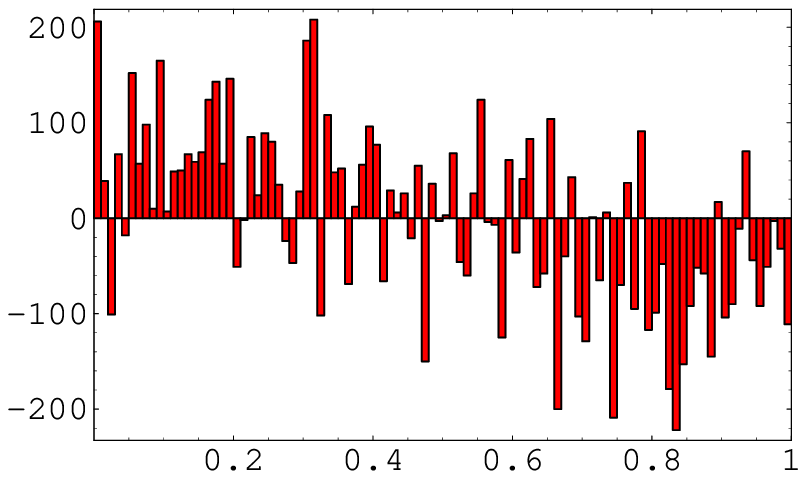}
\includegraphics[width=3.5cm]{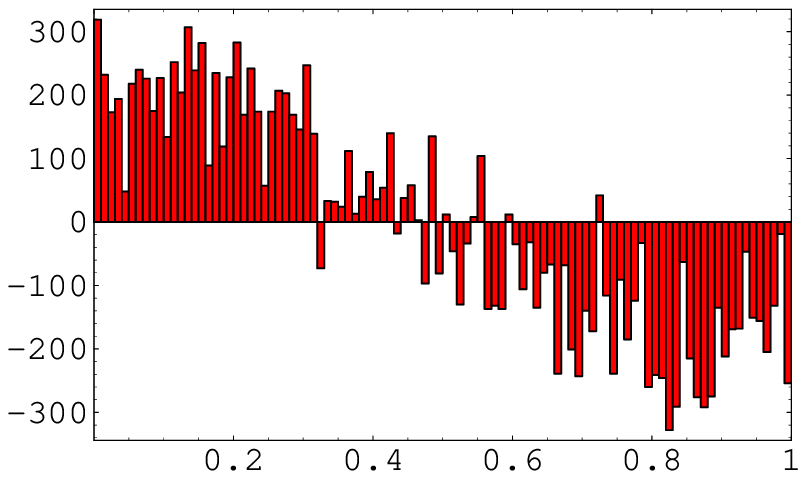}
\includegraphics[width=3.5cm]{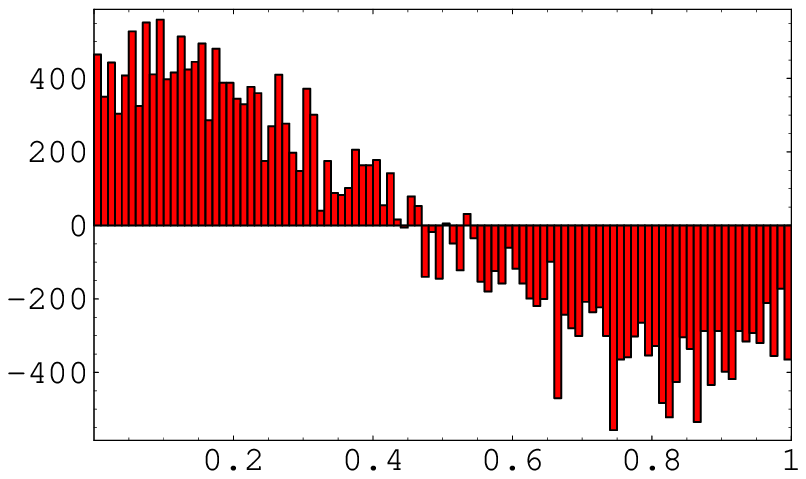}

\includegraphics[width=3.5cm]{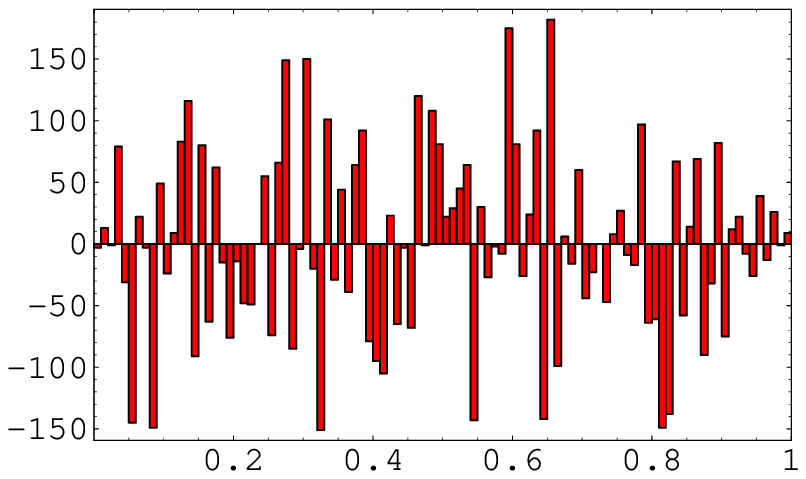}
\includegraphics[width=3.5cm]{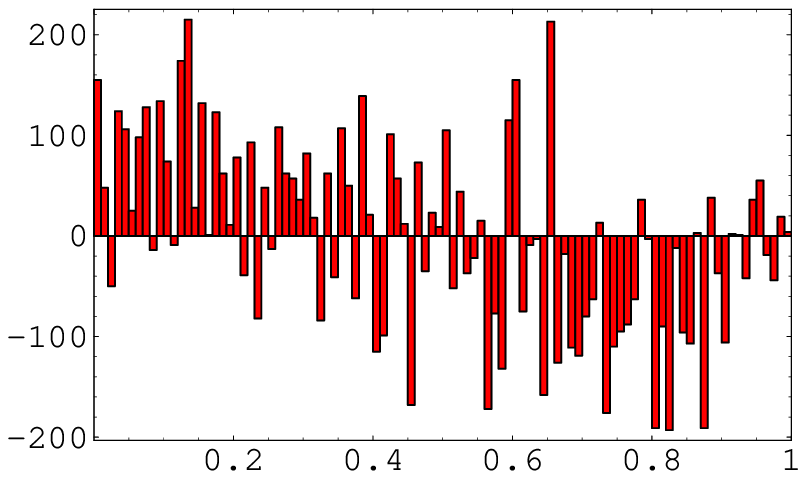}
\includegraphics[width=3.5cm]{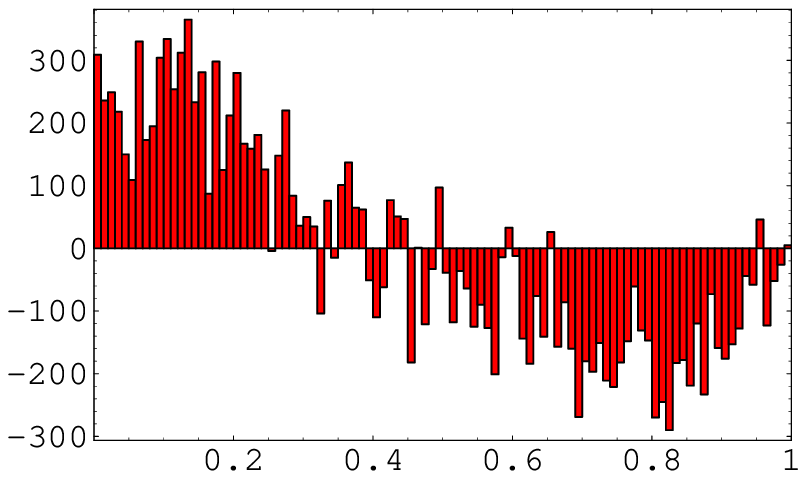}
\includegraphics[width=3.5cm]{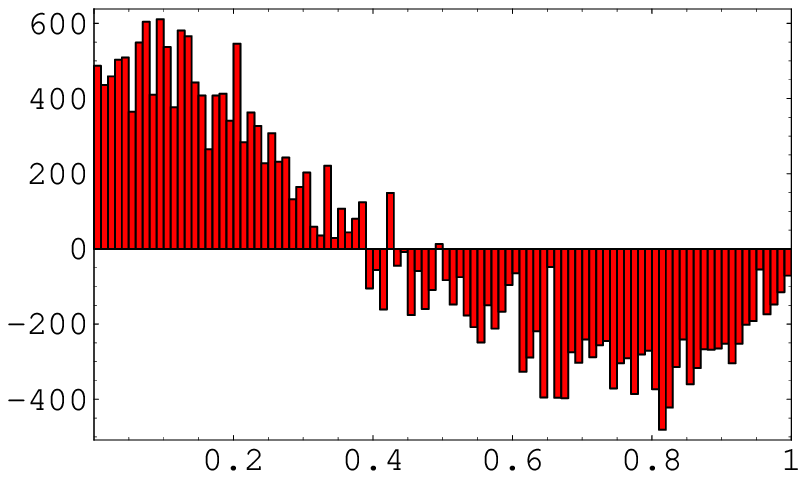}

\includegraphics[width=3.5cm]{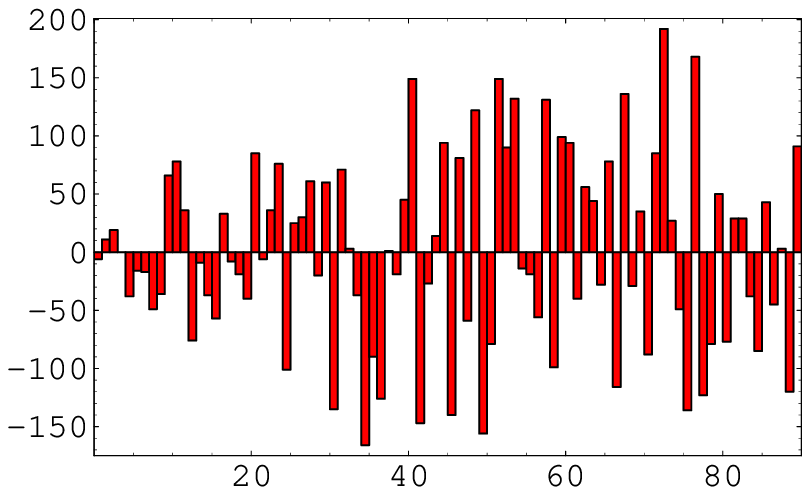}
\includegraphics[width=3.5cm]{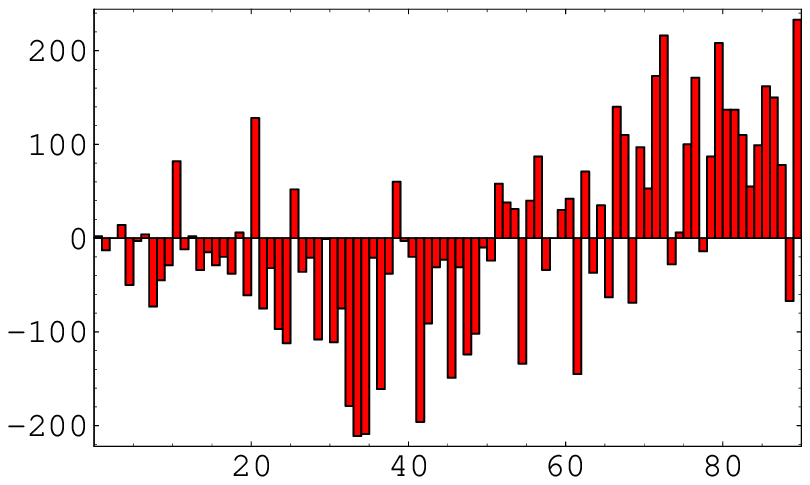}
\includegraphics[width=3.5cm]{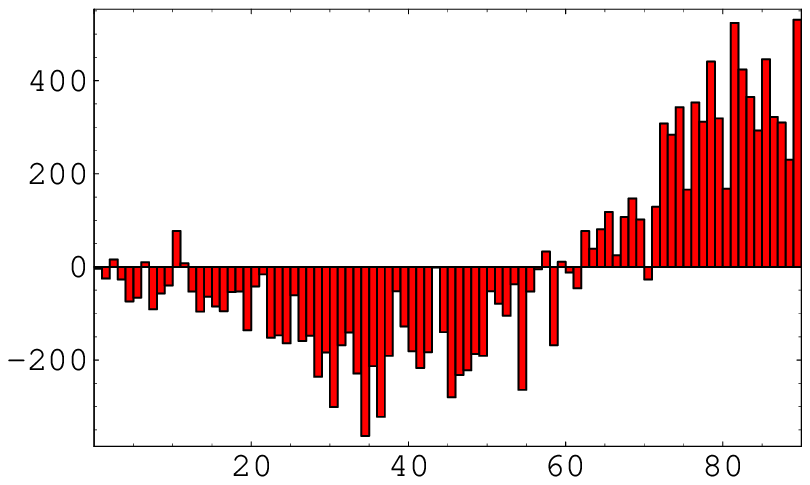}
\includegraphics[width=3.5cm]{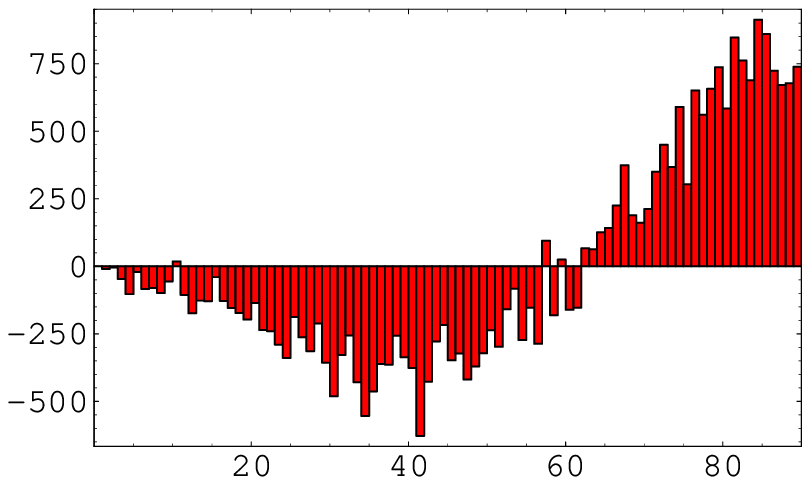}

\caption{Differences between the DSC contaminated distribution and 
the uncontaminated one in $\Lambda$CDM model for the alignment Quadrupole-Dipole
in the case of small $\alpha$, vanishing $\beta$ and nominal scanning strategy. 
First row: $\hat W2 $. Second row: $ W2 $. Third row: $ W2 ^{\circ}$. 
From left to right (in every row) $p=1/1000$, $p=4/1000$, $p=7/1000$, $p=10/1000$.
All the panels present the counts ($y$-axis) versus the statistic ($x$-axis).
See also the text.}

\label{fig2}

\end{figure*}

\begin{figure*}

\centering

\includegraphics[width=3.5cm]{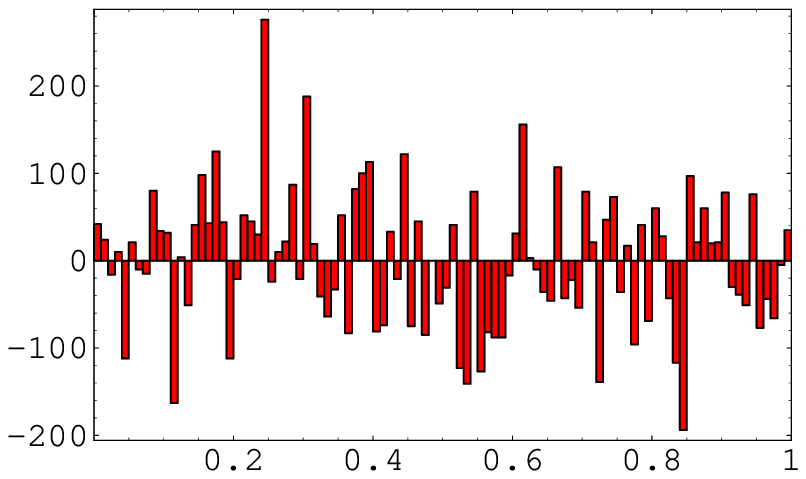}
\includegraphics[width=3.5cm]{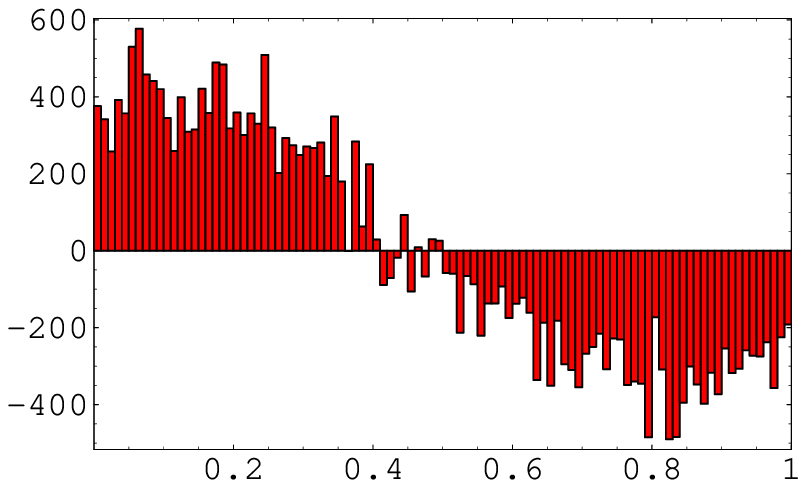}
\includegraphics[width=3.5cm]{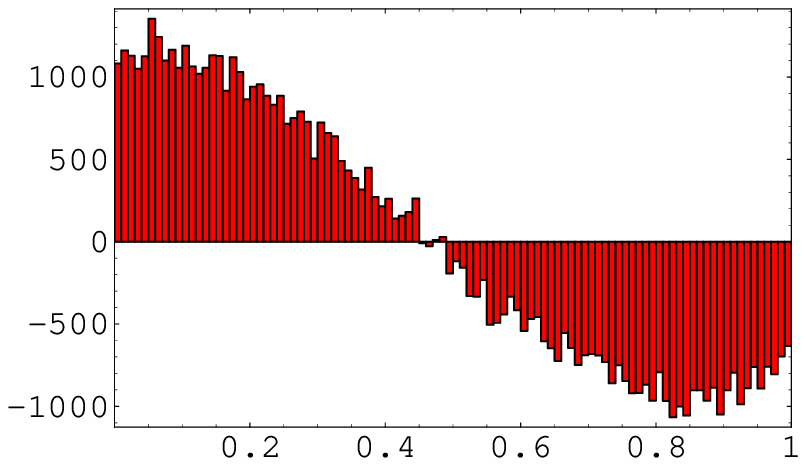}
\includegraphics[width=3.5cm]{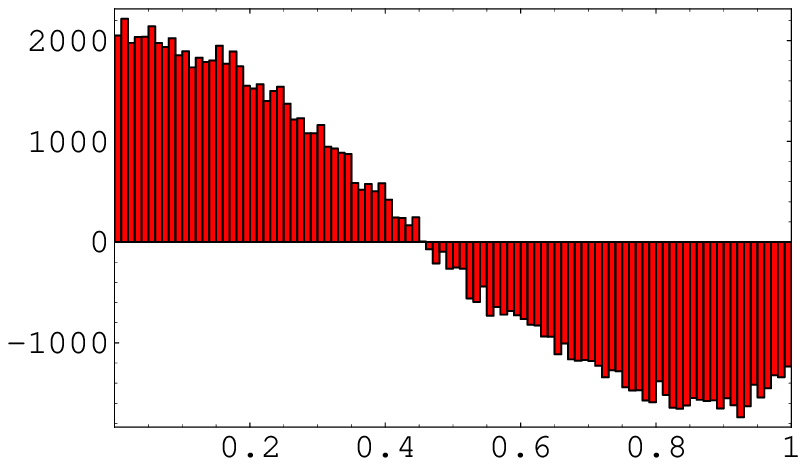}

\includegraphics[width=3.5cm]{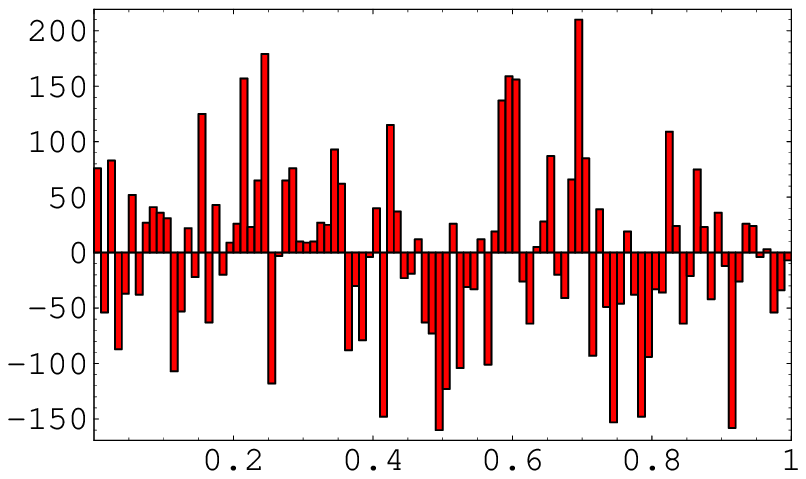}
\includegraphics[width=3.5cm]{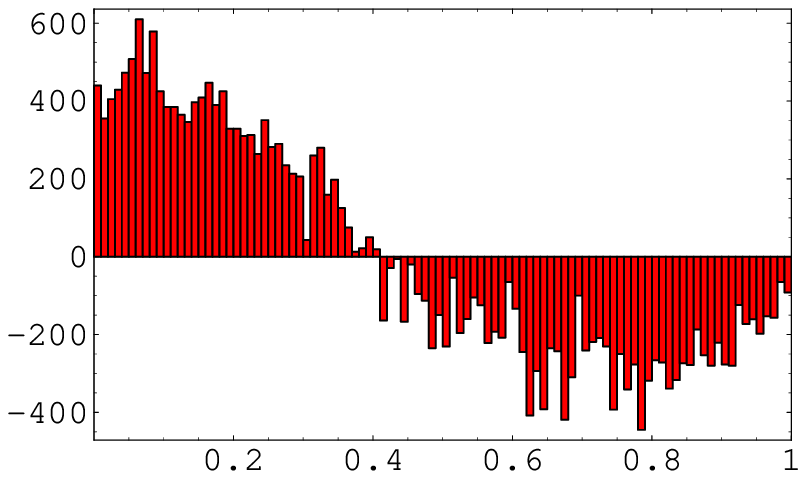}
\includegraphics[width=3.5cm]{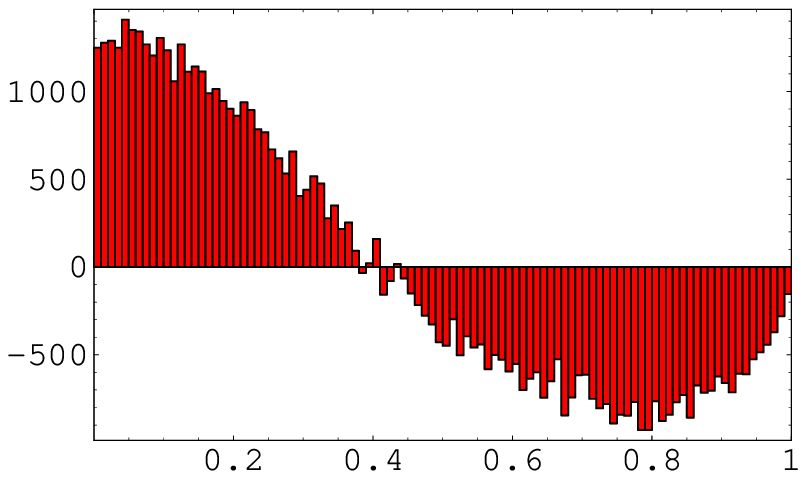}
\includegraphics[width=3.5cm]{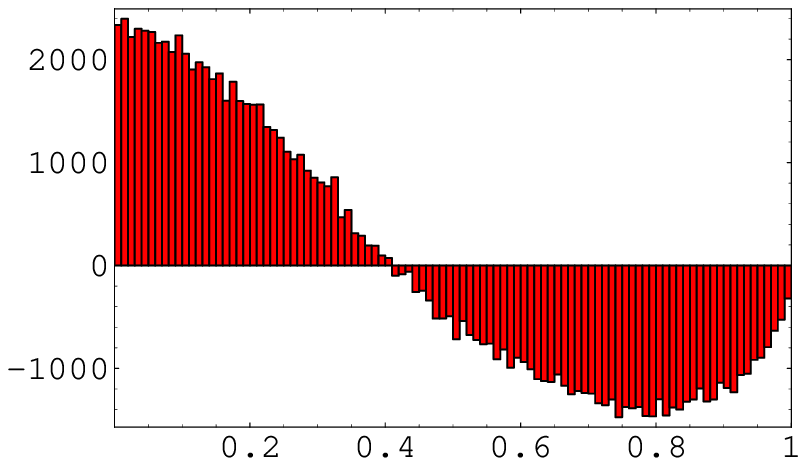}

\includegraphics[width=3.5cm]{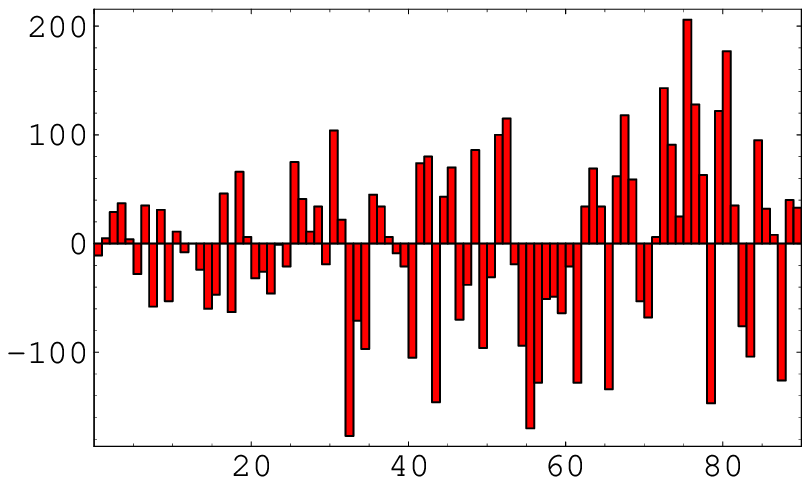}
\includegraphics[width=3.5cm]{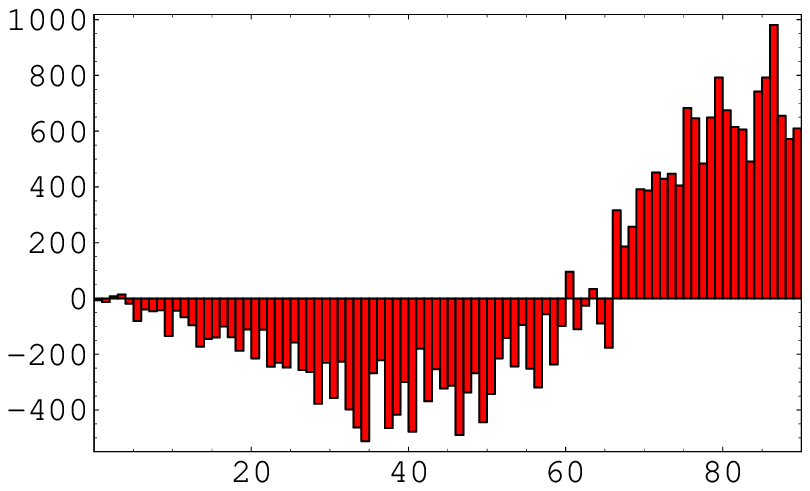}
\includegraphics[width=3.5cm]{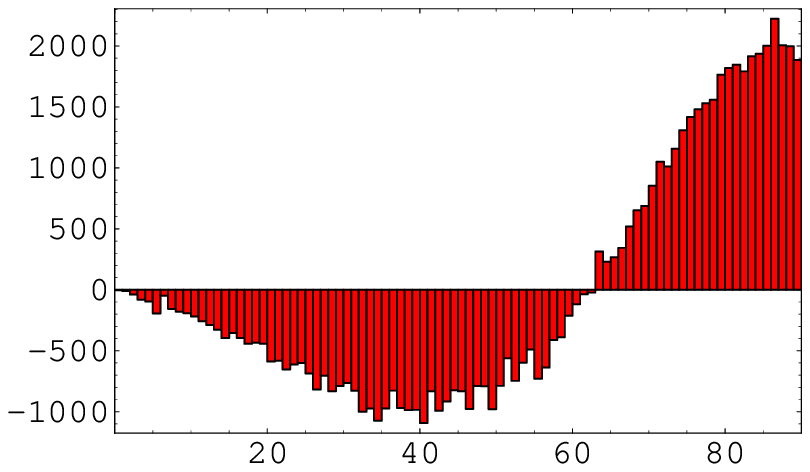}
\includegraphics[width=3.5cm]{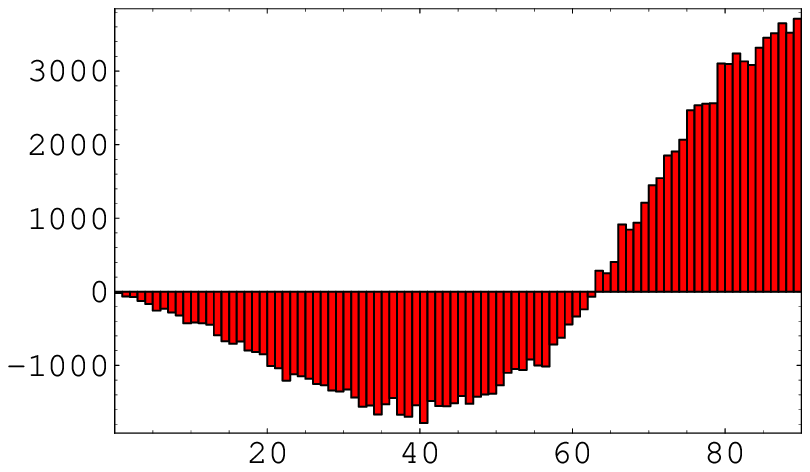}

\caption{Differences between the DSC contaminated distribution and 
the uncontaminated one for the WMAP amplitude for the alignment Quadrupole-Dipole
in the case of small $\alpha$, vanishing $\beta$ and nominal scanning strategy. 
First row: $\hat W2 $. Second row: $ W2 $. Third row: $ W2 ^{\circ}$. 
From left to right (in every row) $p=1/1000$, $p=4/1000$, $p=7/1000$, $p=10/1000$.
All the panels present the counts ($y$-axis) versus the statistic ($x$-axis).
See also the text.}

\label{fig3}

\end{figure*}

\begin{figure*}

\centering

\includegraphics[width=3.5cm]{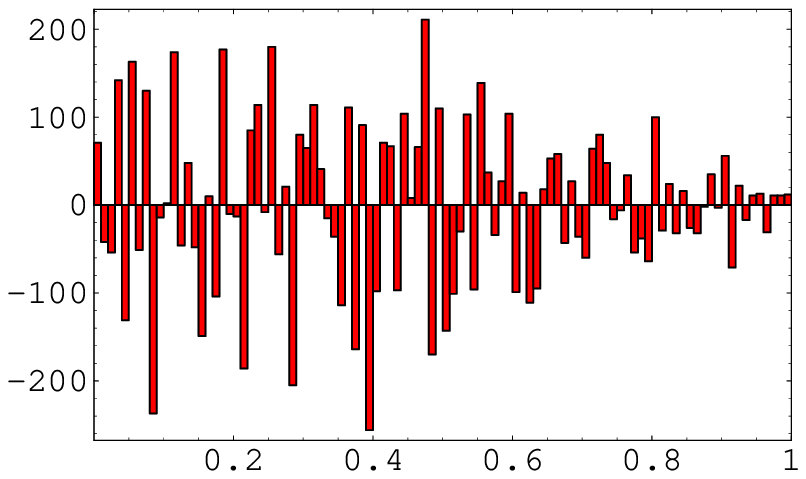}
\includegraphics[width=3.5cm]{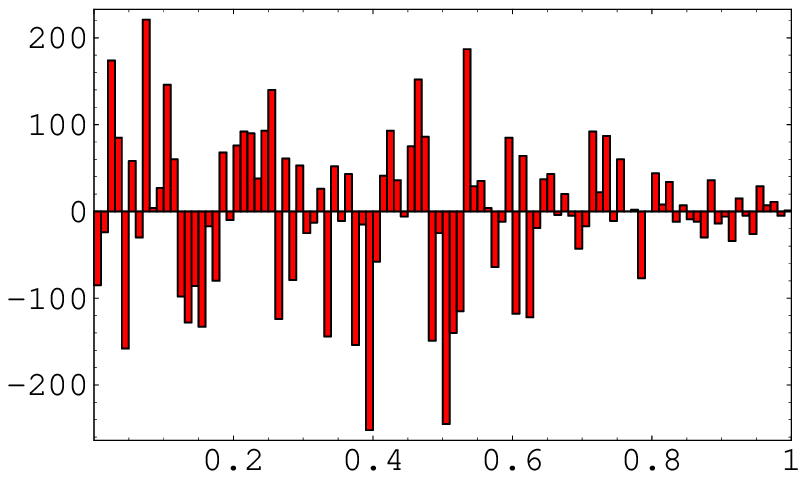}
\includegraphics[width=3.5cm]{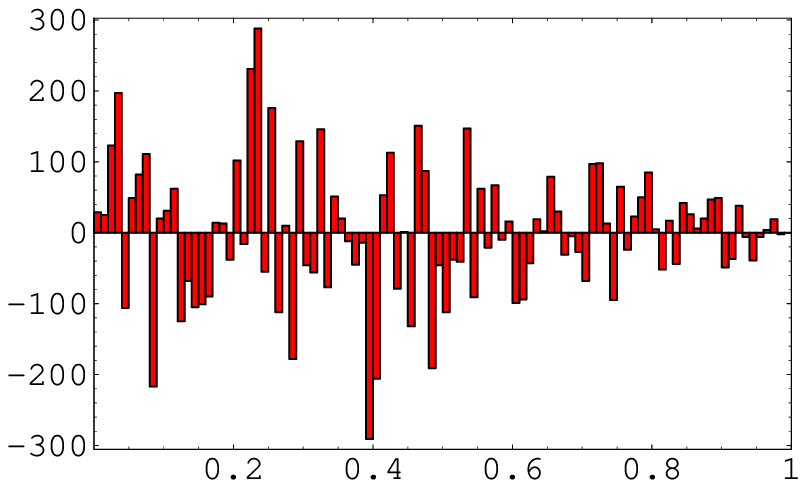}
\includegraphics[width=3.5cm]{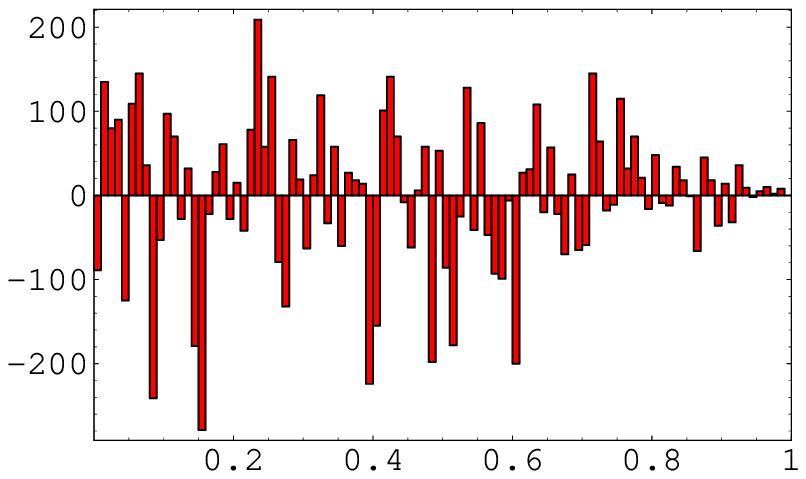}

\includegraphics[width=3.5cm]{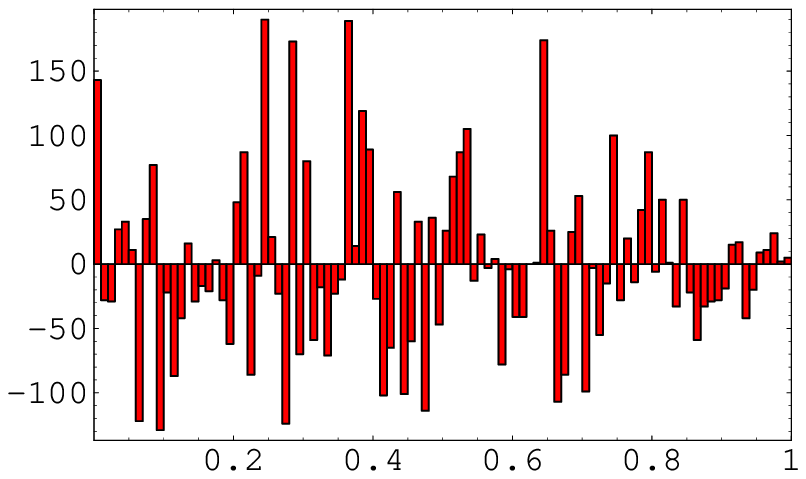}
\includegraphics[width=3.5cm]{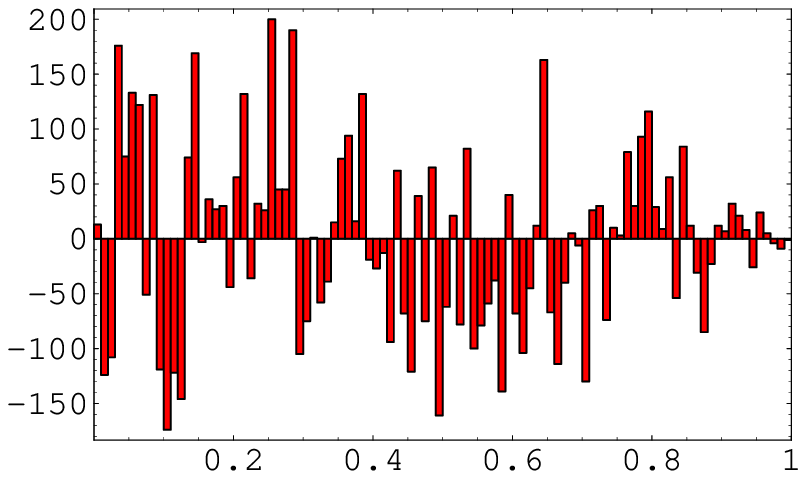}
\includegraphics[width=3.5cm]{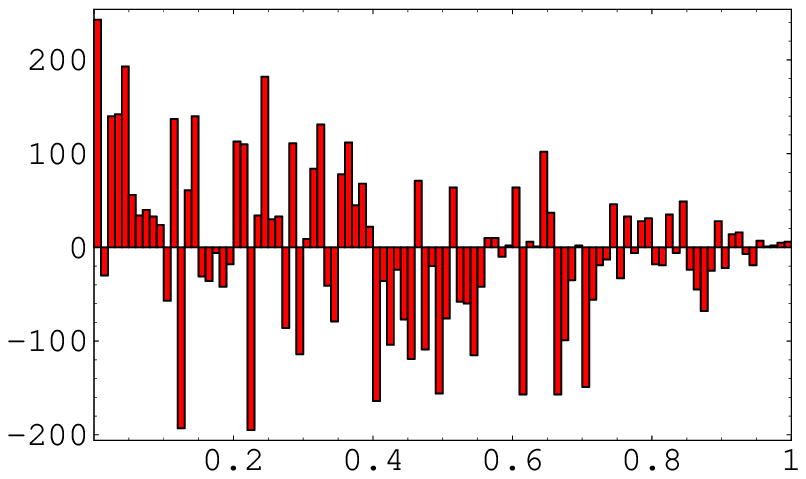}
\includegraphics[width=3.5cm]{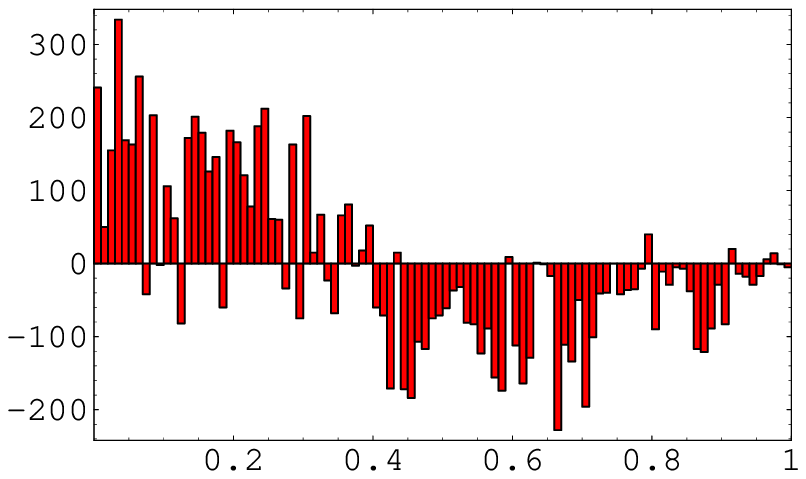}

\caption{Differences between the DSC contaminated distribution and 
the uncontaminated one for the self-alignment Quadrupole-Quadrupole 
in the case of small $\alpha$, vanishing $\beta$ and nominal scanning strategy. 
First row: $R22$ in $\Lambda$CDM model. Second row: $R22$ for the WMAP amplitude of the 
intrinsic sky. From left to right (in every row) $p=1/1000$, 
$p=4/1000$, $p=7/1000$, $p=10/1000$. All the panels present the counts ($y$-axis) versus the statistic ($x$-axis).
See also the text.}

\label{fig4}

\end{figure*}

\begin{figure*}

\centering

\includegraphics[width=3.5cm]{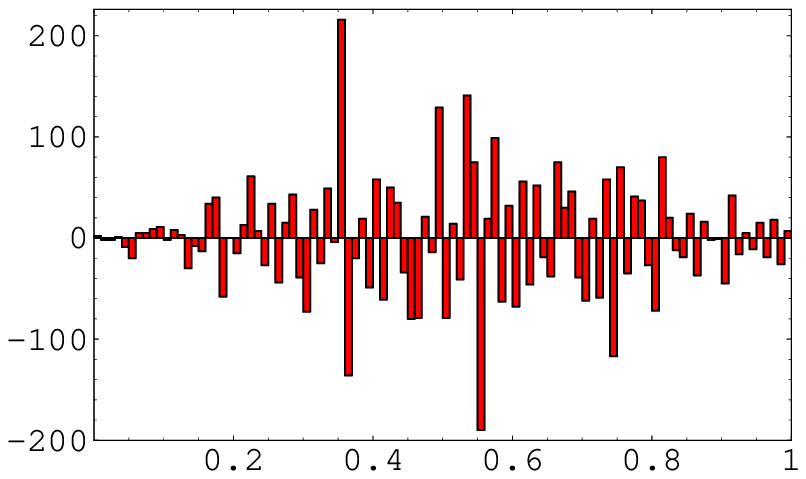}
\includegraphics[width=3.5cm]{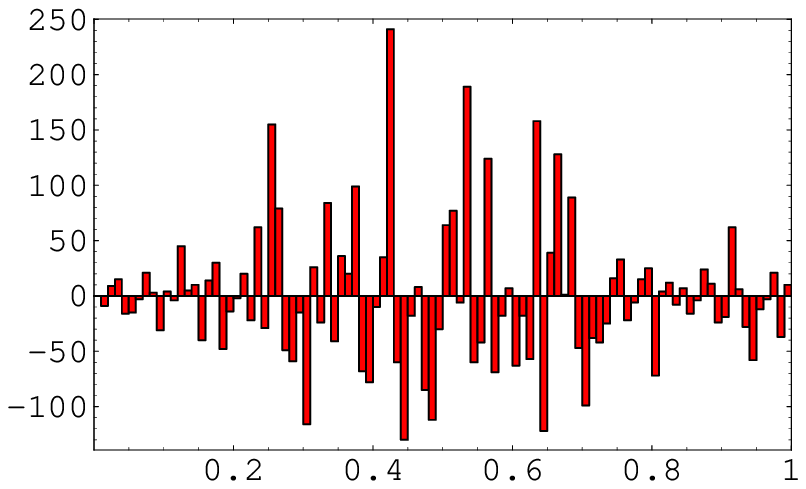}
\includegraphics[width=3.5cm]{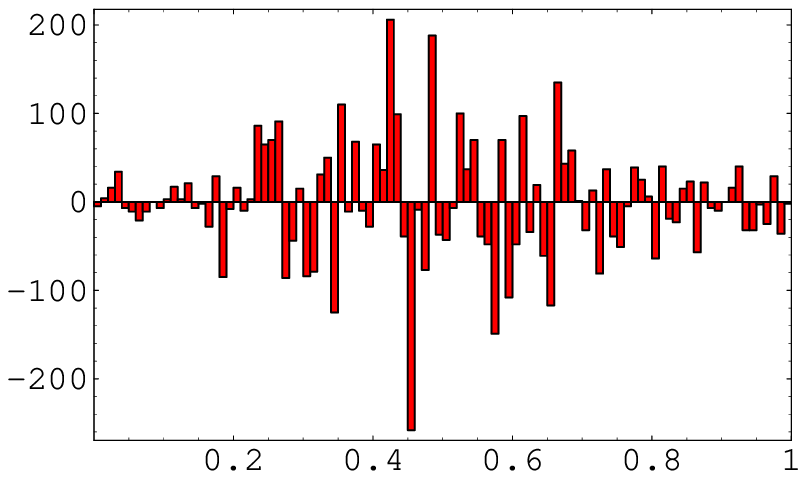}
\includegraphics[width=3.5cm]{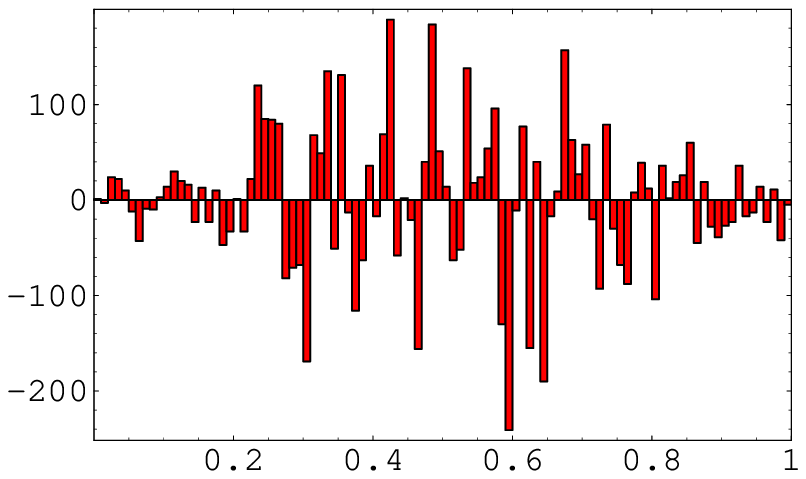}

\includegraphics[width=3.5cm]{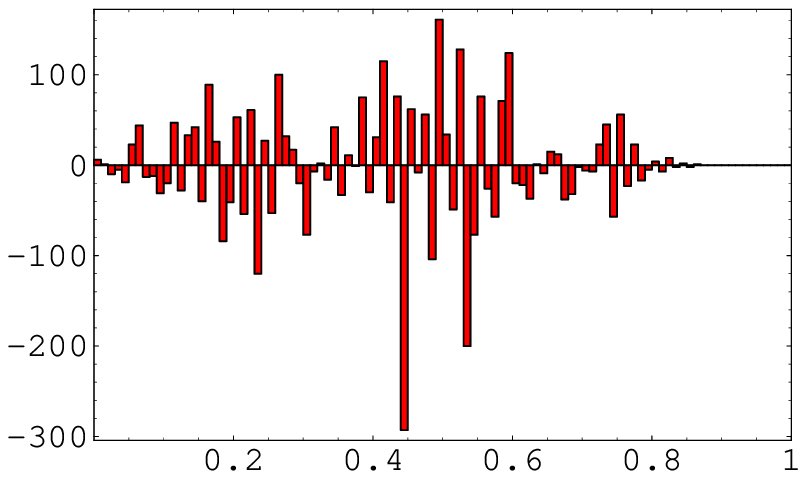}
\includegraphics[width=3.5cm]{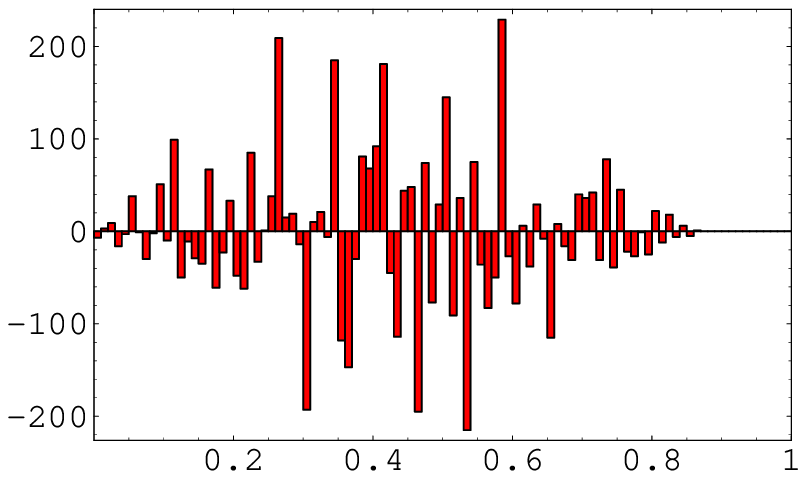}
\includegraphics[width=3.5cm]{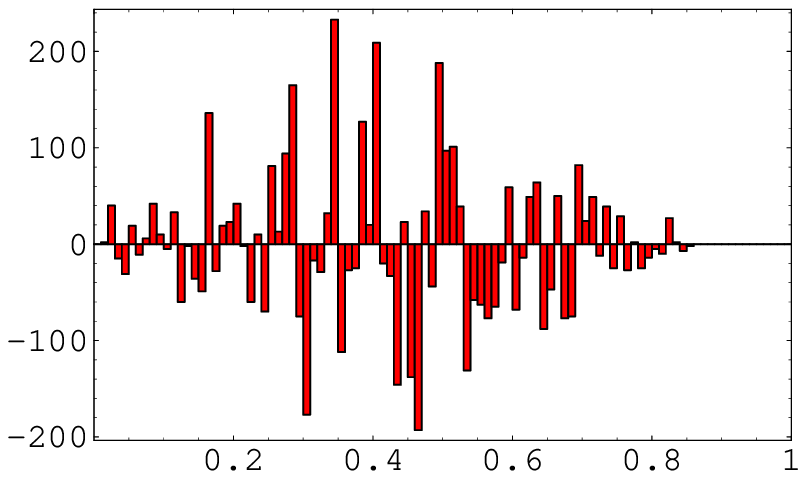}
\includegraphics[width=3.5cm]{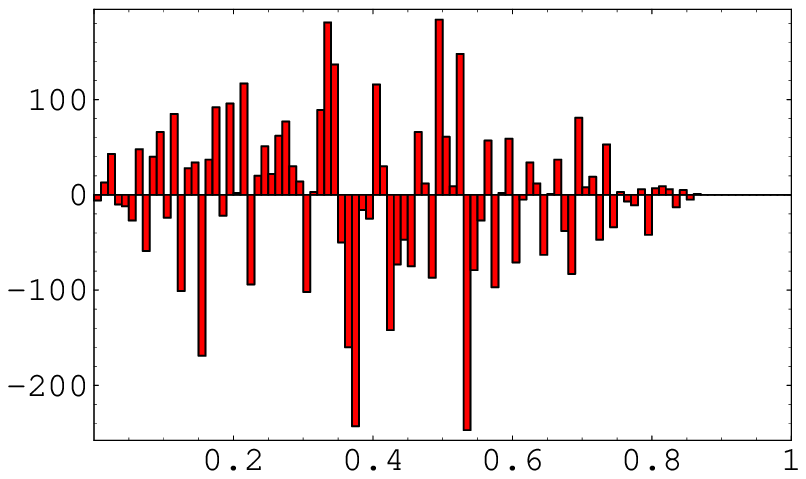}

\caption{Differences between the DSC contaminated distribution and 
the uncontaminated one in $\Lambda$CDM model for the alignment Octupole-Quadrupole 
in the case of small $\alpha$, vanishing $\beta$ and nominal scanning strategy. 
First row: $D23$. Second row: $S23$. From left to right (in every row) $p=1/1000$, 
$p=4/1000$, $p=7/1000$, $p=10/1000$. All the panels present the counts ($y$-axis) versus the statistic ($x$-axis).
See also the text.}

\label{fig5}

\end{figure*}

\begin{figure*}

\centering

\includegraphics[width=3.5cm]{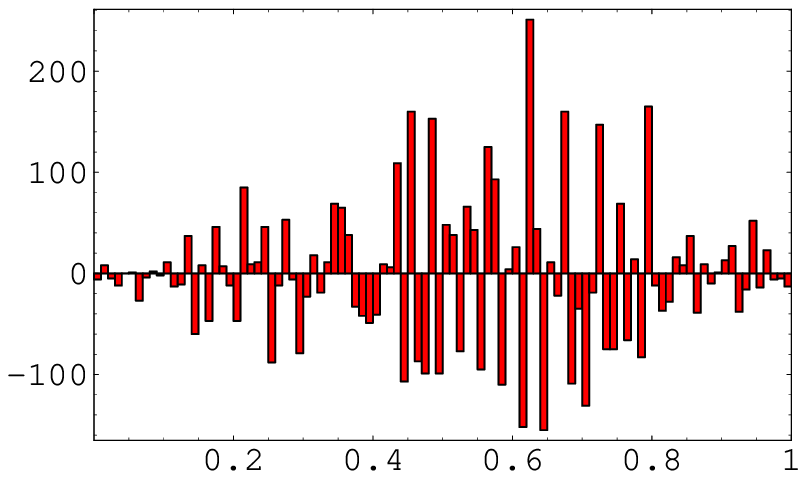}
\includegraphics[width=3.5cm]{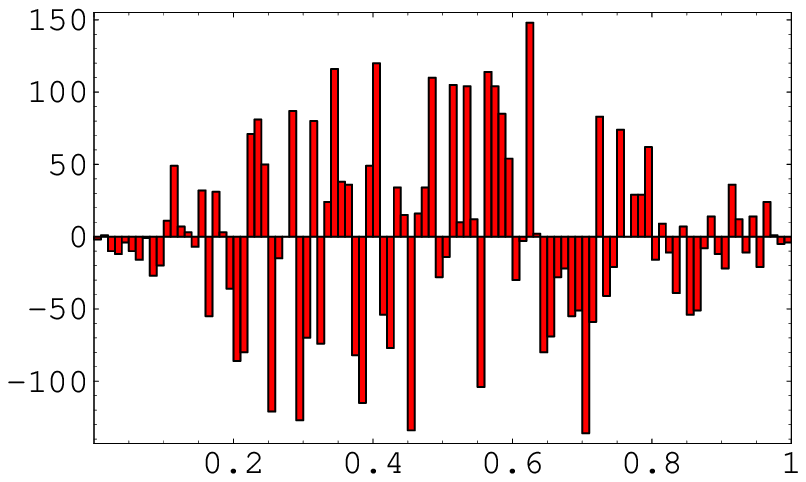}
\includegraphics[width=3.5cm]{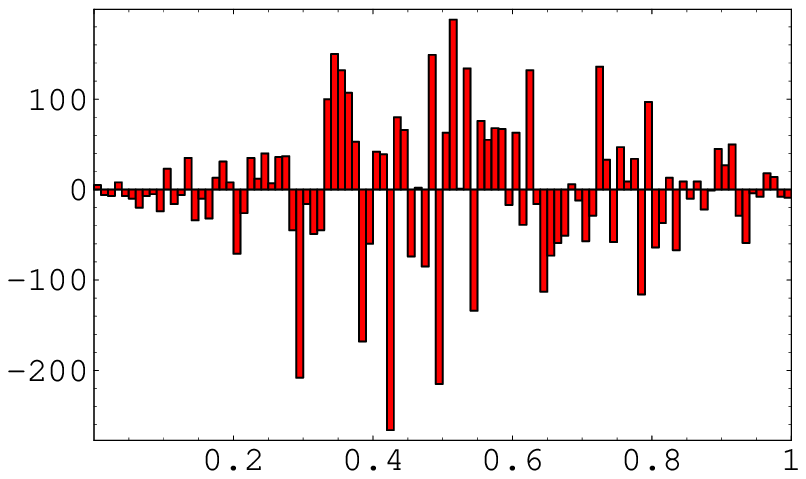}
\includegraphics[width=3.5cm]{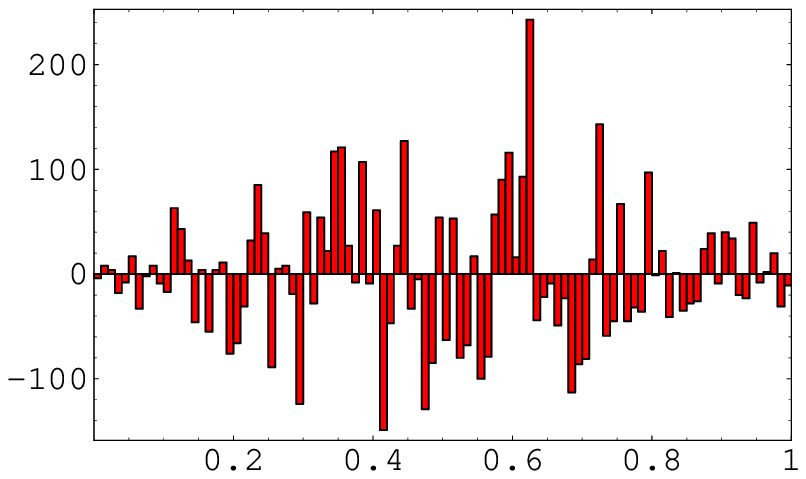}

\includegraphics[width=3.5cm]{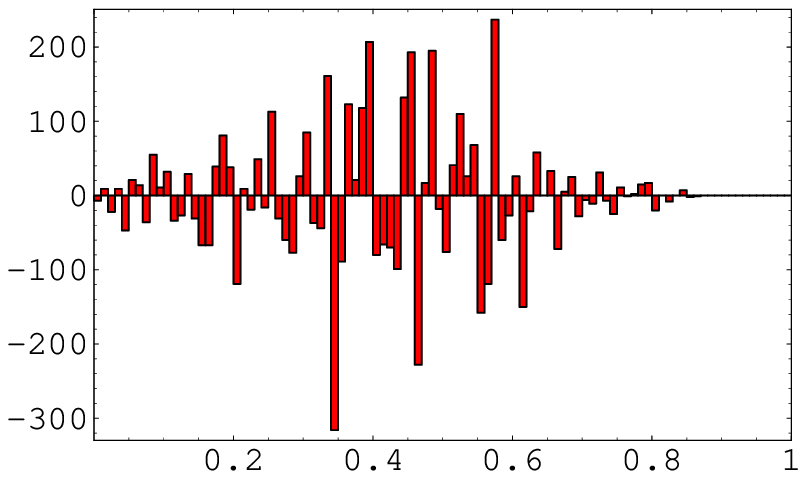}
\includegraphics[width=3.5cm]{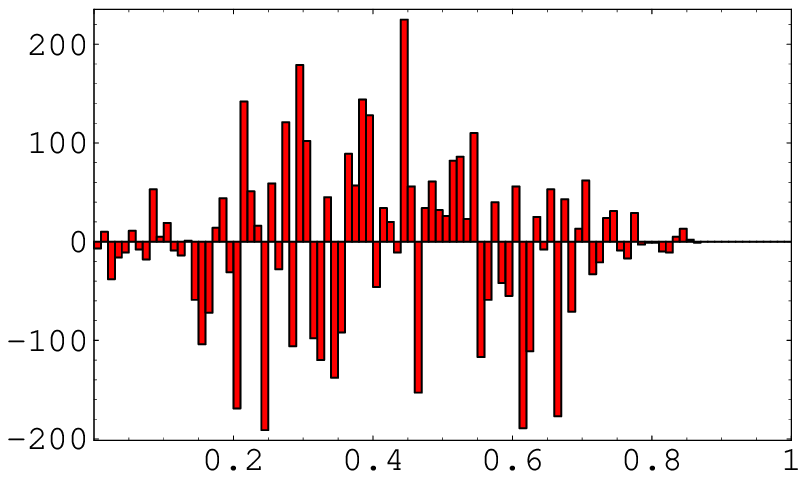}
\includegraphics[width=3.5cm]{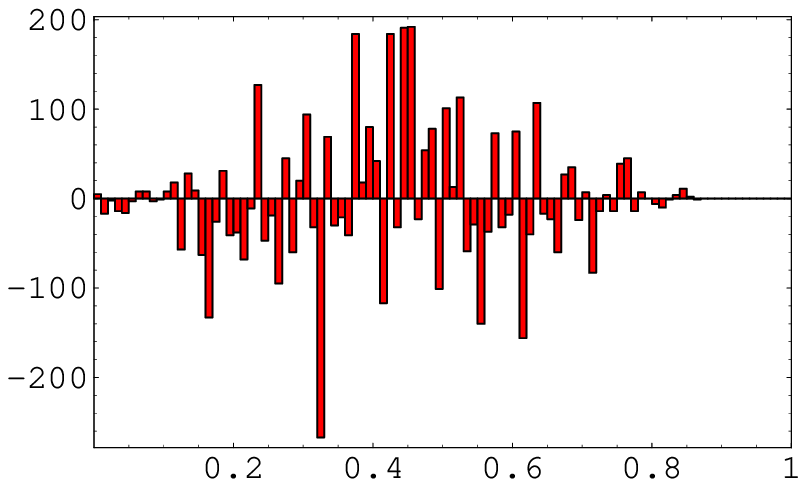}
\includegraphics[width=3.5cm]{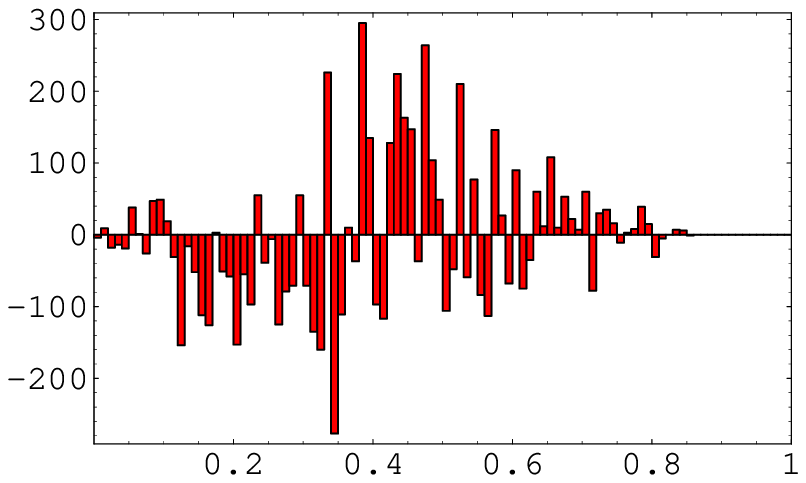}

\caption{Differences between the DSC contaminated distribution and 
the uncontaminated one for the WMAP amplitude for the alignment Octupole-Quadrupole
in the case of small $\alpha$, vanishing $\beta$ and nominal scanning strategy. 
First row: $D23$. Second row: $S23$. 
From left to right (in every row) $p=1/1000$, $p=4/1000$, $p=7/1000$, $p=10/1000$.
All the panels present the counts ($y$-axis) versus the statistic ($x$-axis).
See also the text.}

\label{fig6}
\end{figure*}

\section{Results}
\label{results}

In Fig. \ref{fig1} we plot the uncontaminated distribution
for $\hat W2$, $W2$, $W2^{\circ}$, $R22$, $\hat W3$, $ W3$, $ D23$, $S23$, 
$D33$, $S33$, $\hat W4$, $W4$, $D42$, $S42$, $D43$, $S43$, $D44$, $S44$. 
Notice the perfect agreement between the analytical, see equation~(\ref{anforr22}), and the numerical distribution of $R22$.
For sake of completeness we report in Table \ref{one}
the values of the considered estimators obtained, for example, 
adopting the CMB anisotropy component in the V band of 
the WMAP $3$ year release with a Kp2 mask
and the dipole given in \cite{hinshaw}.

\begin{figure*}

\centering

\includegraphics[width=3.5cm]{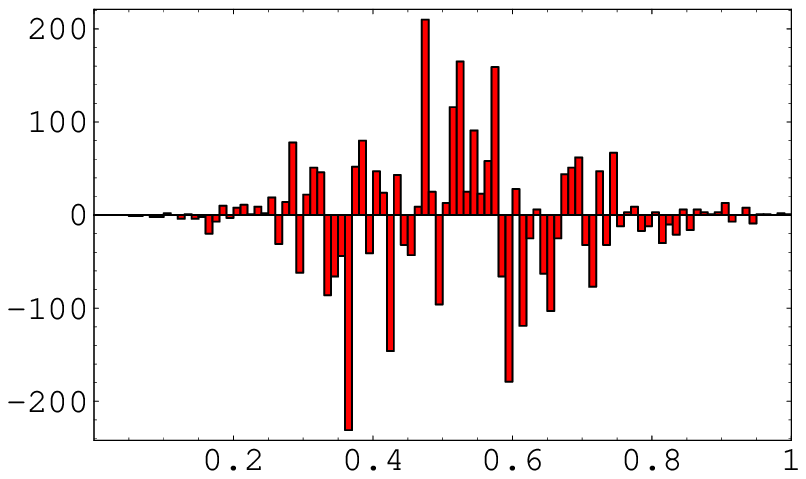}
\includegraphics[width=3.5cm]{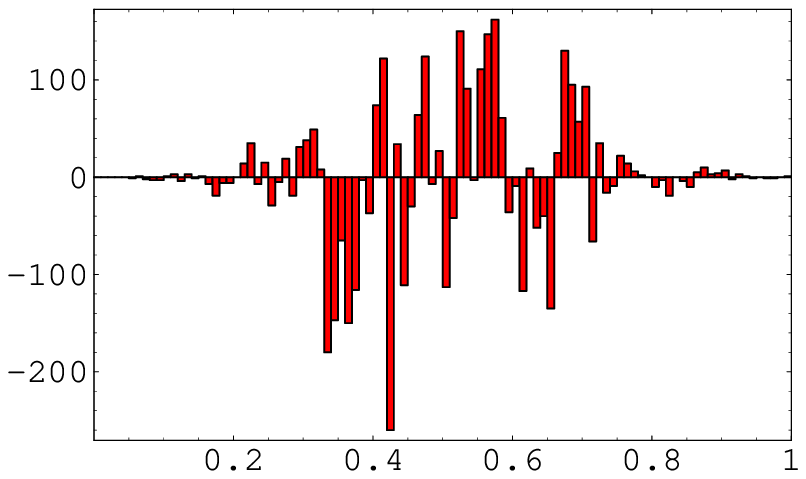}
\includegraphics[width=3.5cm]{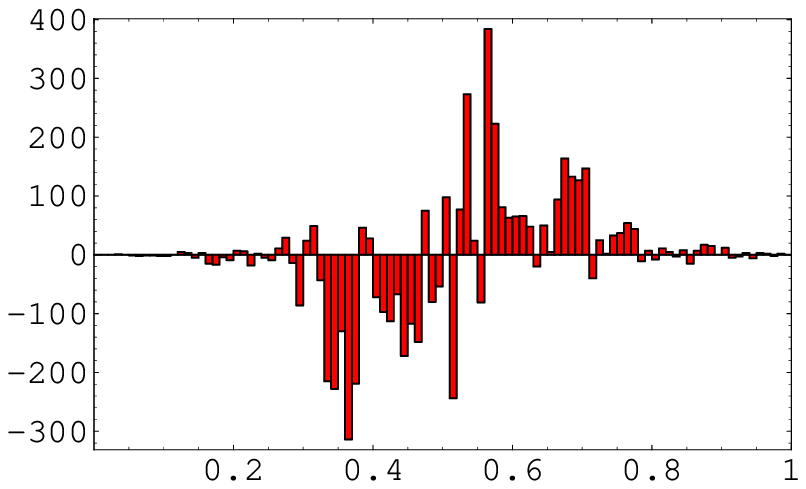}
\includegraphics[width=3.5cm]{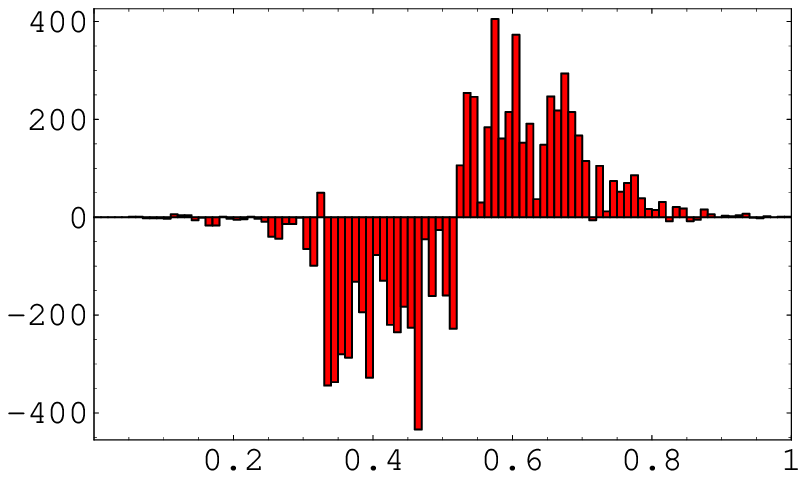}

\includegraphics[width=3.5cm]{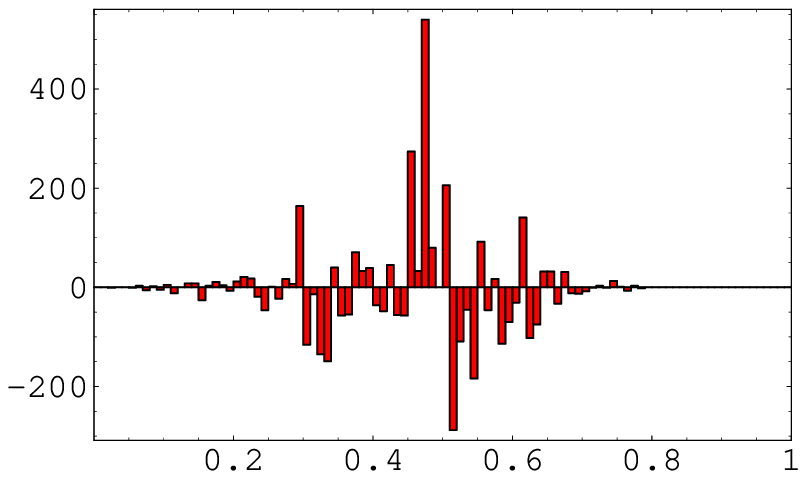}
\includegraphics[width=3.5cm]{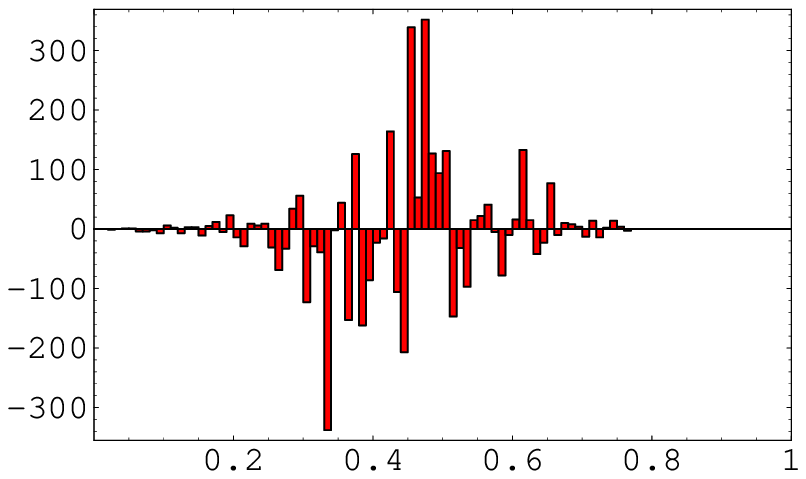}
\includegraphics[width=3.5cm]{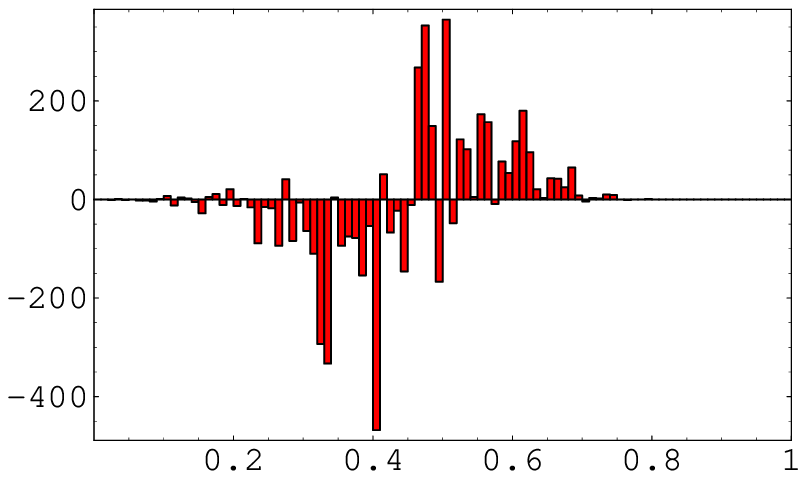}
\includegraphics[width=3.5cm]{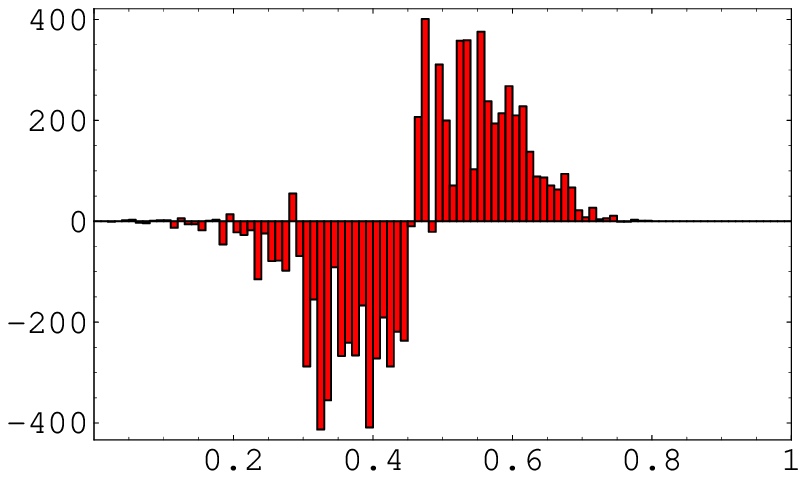}

\caption{Differences between the DSC contaminated distribution and 
the uncontaminated one in $\Lambda$CDM model for the alignment Hexadecapole-Dipole
in the case of small $\alpha$, vanishing $\beta$ and nominal scanning strategy. 
First row: $\hat W4$. Second row: $ W 4$. 
From left to right (in every row) $p=1/1000$, $p=4/1000$, $p=7/1000$, $p=10/1000$.
All the panels present the counts ($y$-axis) versus the statistic ($x$-axis).
See also the text.}

\label{fig7}

\end{figure*}

\begin{figure*}

\centering

\includegraphics[width=3.5cm]{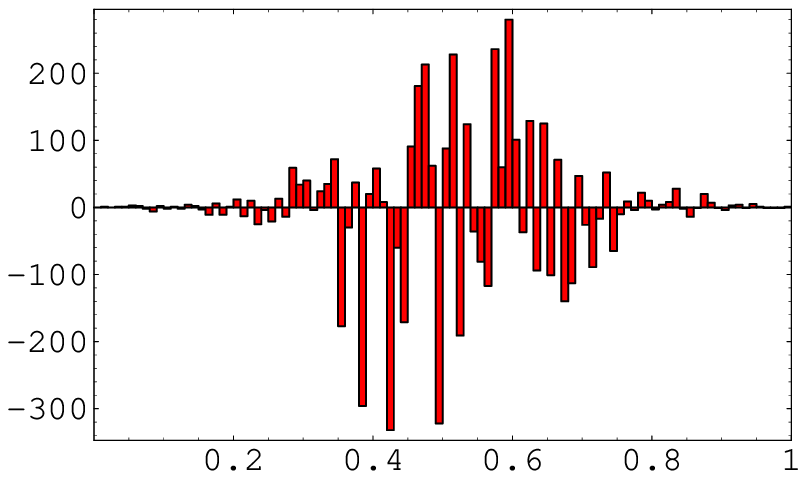}
\includegraphics[width=3.5cm]{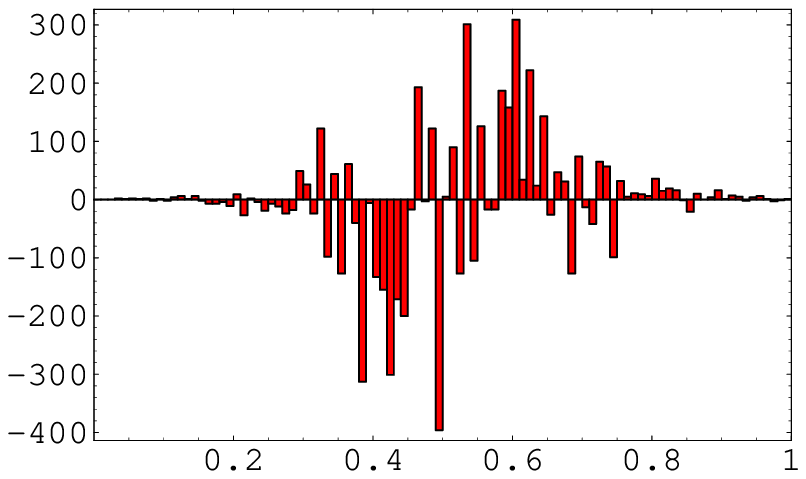}
\includegraphics[width=3.5cm]{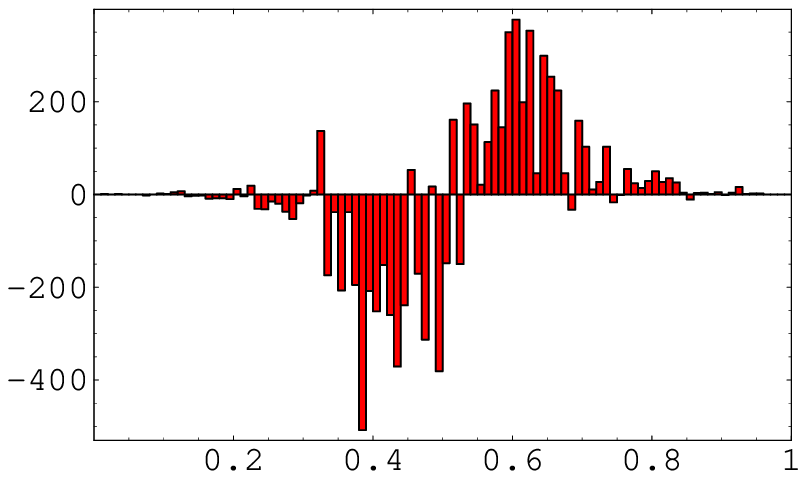}
\includegraphics[width=3.5cm]{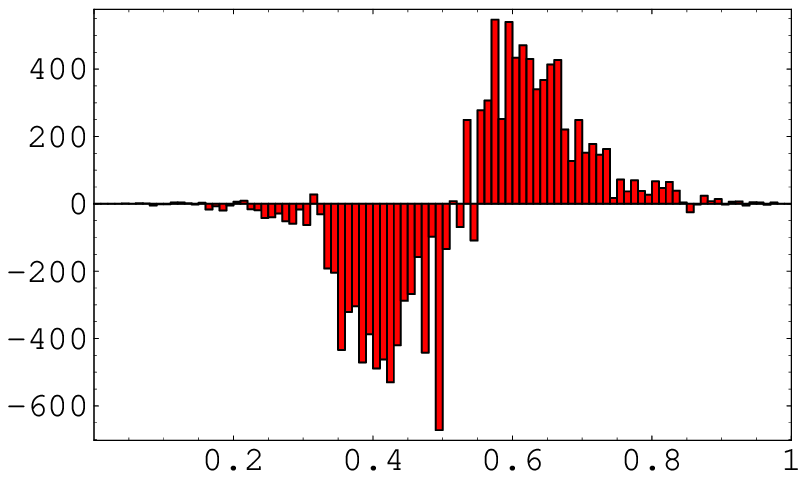}

\includegraphics[width=3.5cm]{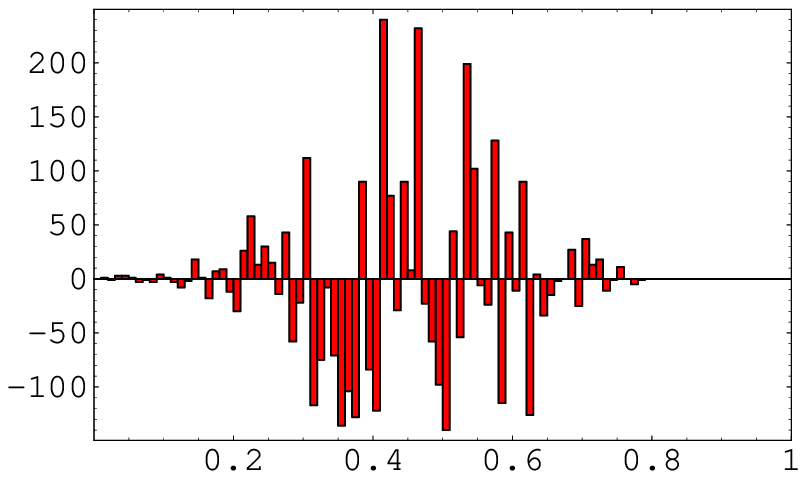}
\includegraphics[width=3.5cm]{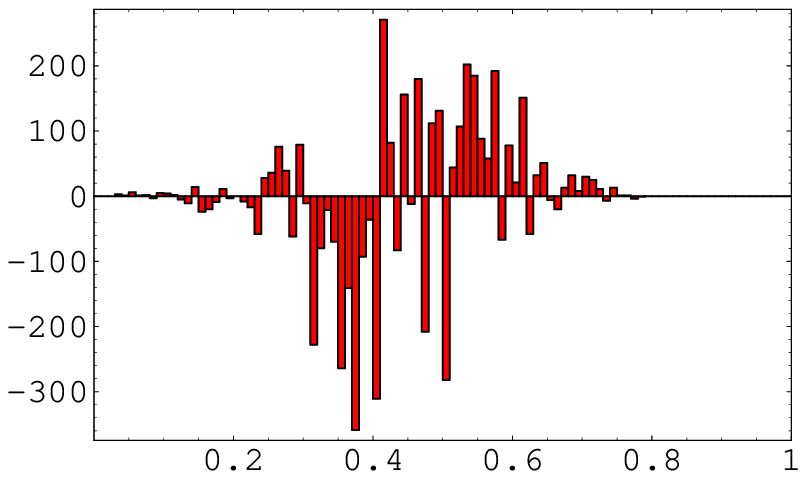}
\includegraphics[width=3.5cm]{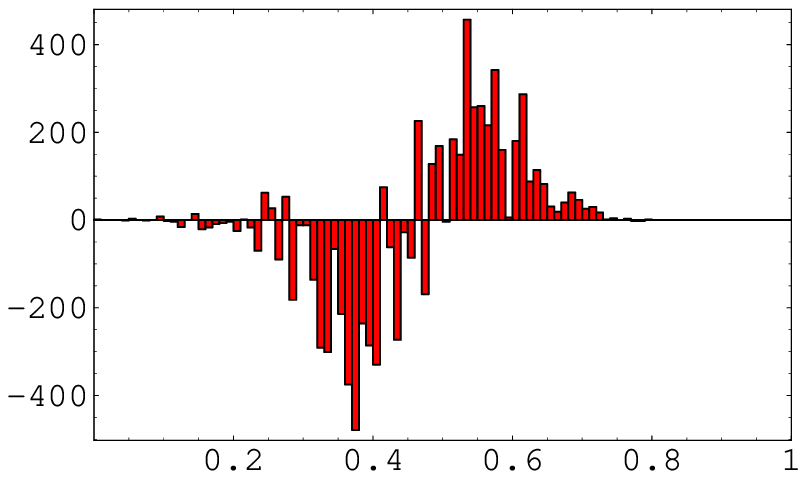}
\includegraphics[width=3.5cm]{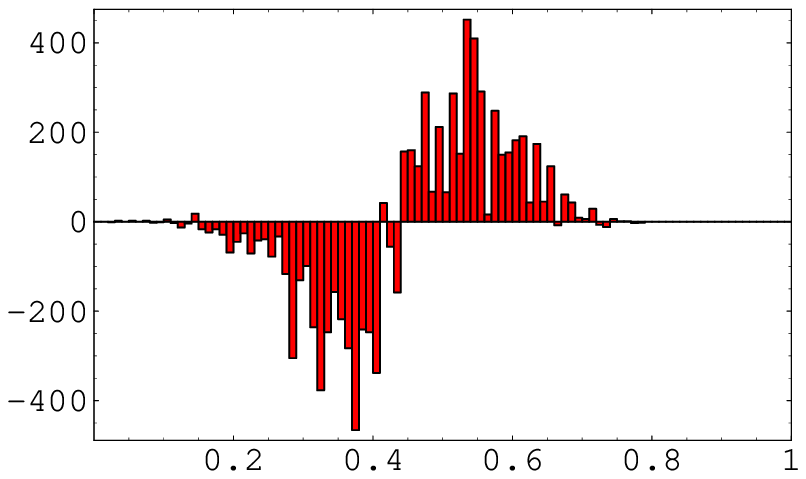}

\caption{Differences between the DSC contaminated distribution and 
the uncontaminated one for the WMAP amplitude for the alignment Hexadecapole-Dipole
in the case of small $\alpha$, vanishing $\beta$ and nominal scanning strategy. 
First row: $\hat W4$. Second row: $ W 4$. 
From left to right (in every row) $p=1/1000$, $p=4/1000$, $p=7/1000$, $p=10/1000$.
All the panels present the counts ($y$-axis) versus the statistic ($x$-axis).
See also the text.}

\label{fig8}
\end{figure*}

\begin{figure*}

\centering

\includegraphics[width=3.5cm]{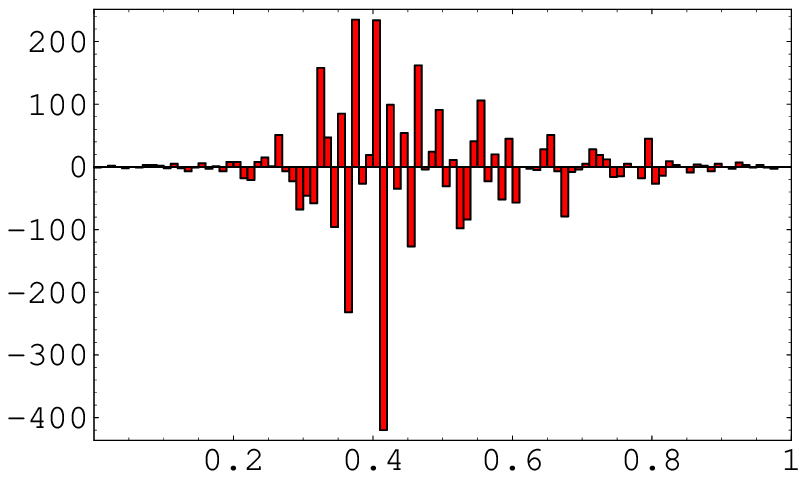}
\includegraphics[width=3.5cm]{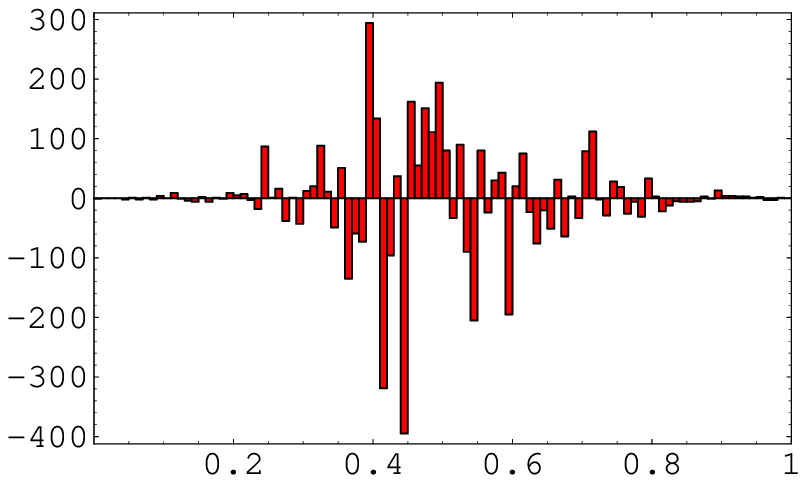}
\includegraphics[width=3.5cm]{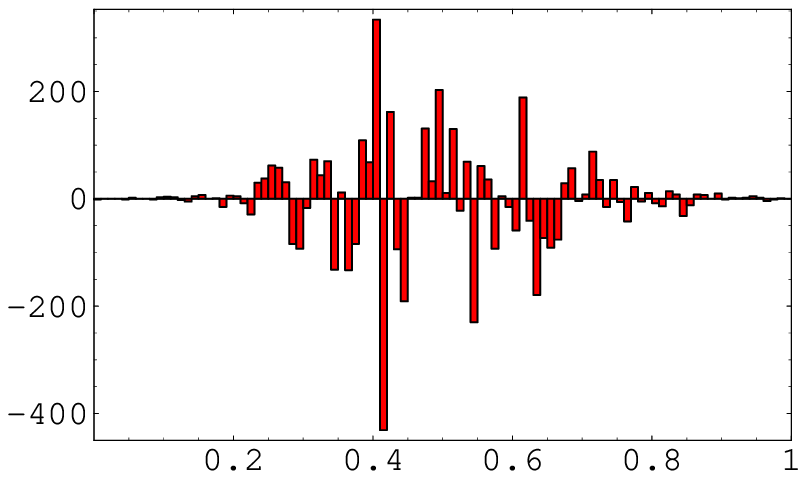}
\includegraphics[width=3.5cm]{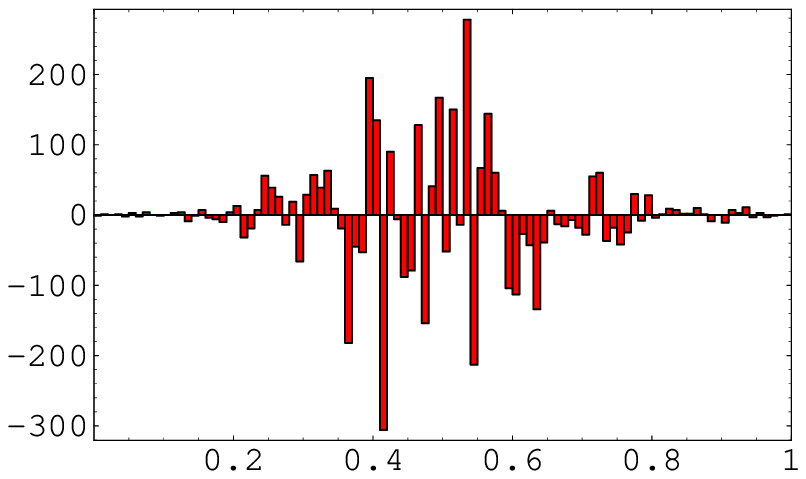}

\includegraphics[width=3.5cm]{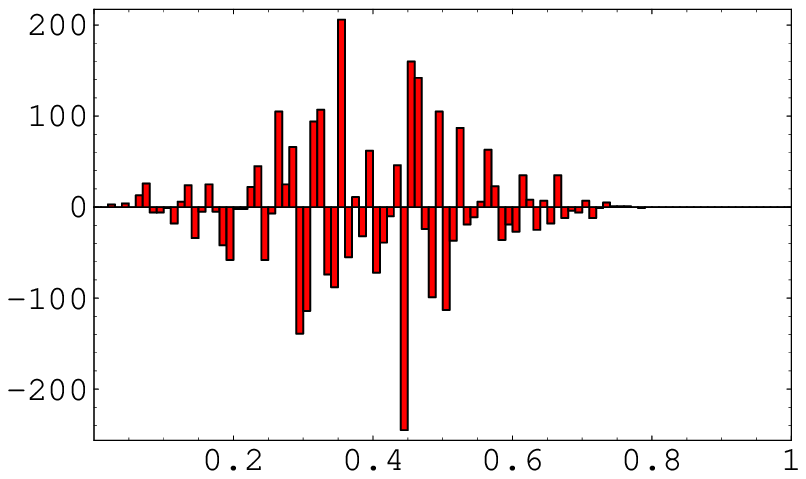}
\includegraphics[width=3.5cm]{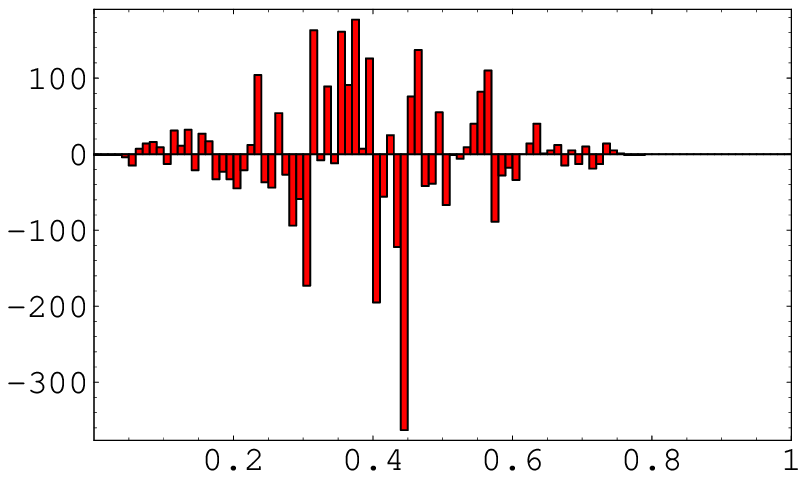}
\includegraphics[width=3.5cm]{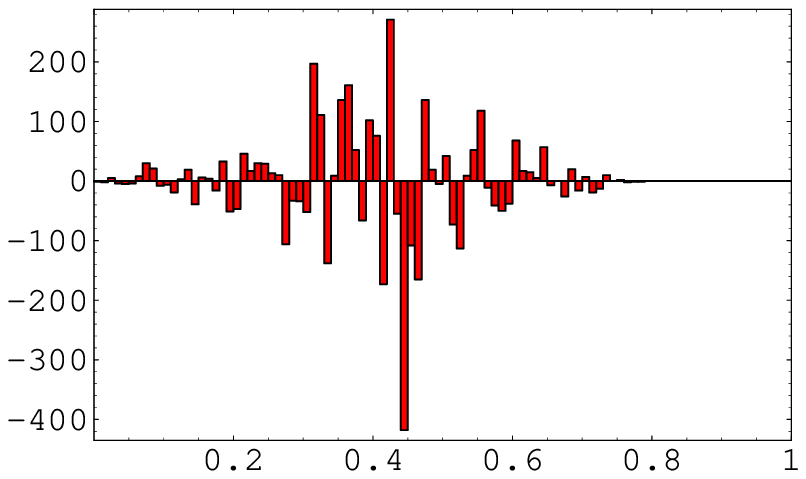}
\includegraphics[width=3.5cm]{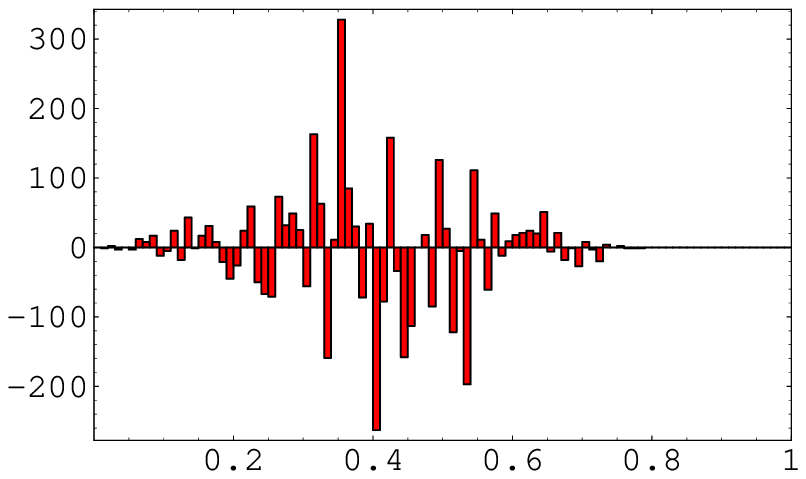}

\caption{Differences between the DSC contaminated distribution and 
the uncontaminated one in $\Lambda$CDM model for the alignment Hexadecapole-Quadrupole
in the case of small $\alpha$, vanishing $\beta$ and nominal scanning strategy. 
First row: $D42$. Second row: $S42$. 
From left to right (in every row) $p=1/1000$, $p=4/1000$, $p=7/1000$, $p=10/1000$.
All the panels present the counts ($y$-axis) versus the statistic ($x$-axis).
See also the text.}

\label{fig9}

\end{figure*}

\begin{figure*}

\centering

\includegraphics[width=3.5cm]{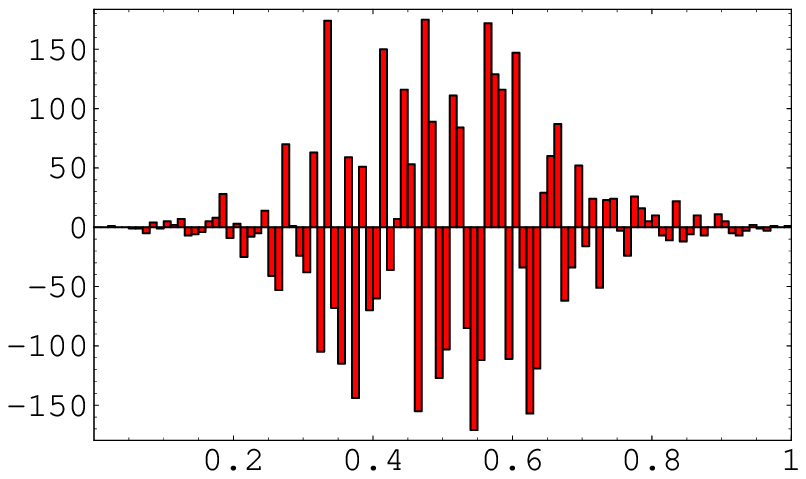}
\includegraphics[width=3.5cm]{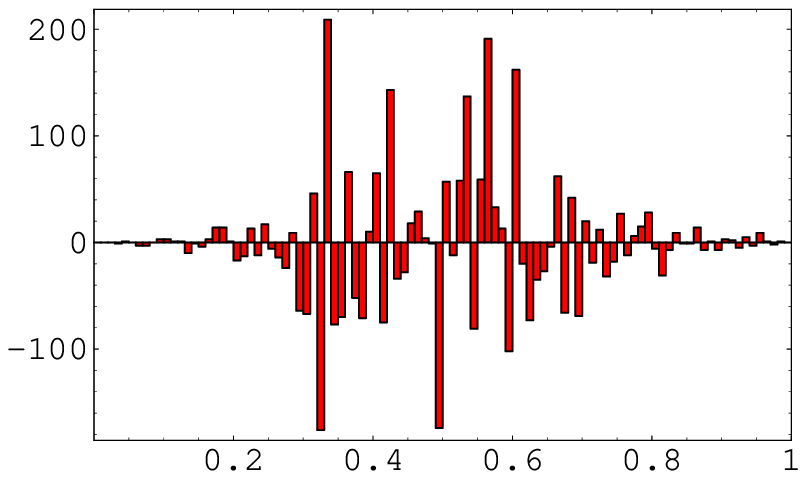}
\includegraphics[width=3.5cm]{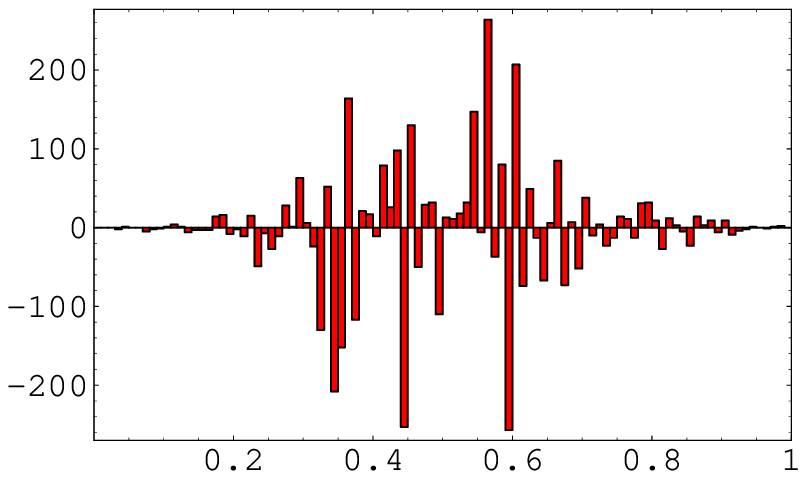}
\includegraphics[width=3.5cm]{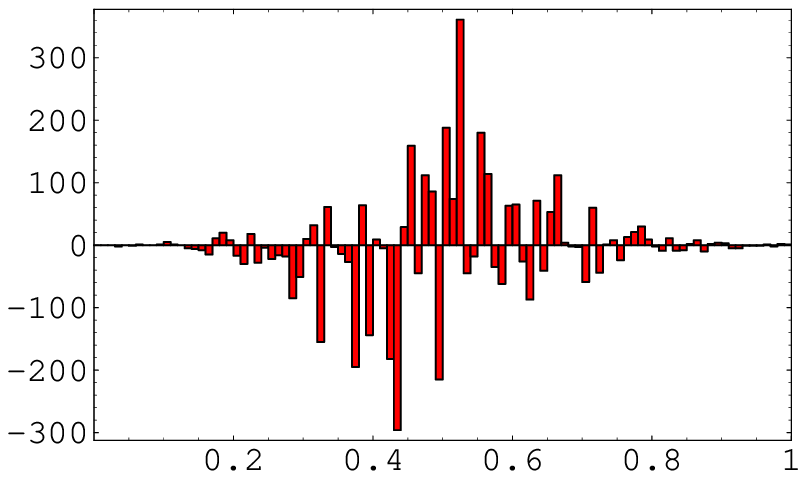}

\includegraphics[width=3.5cm]{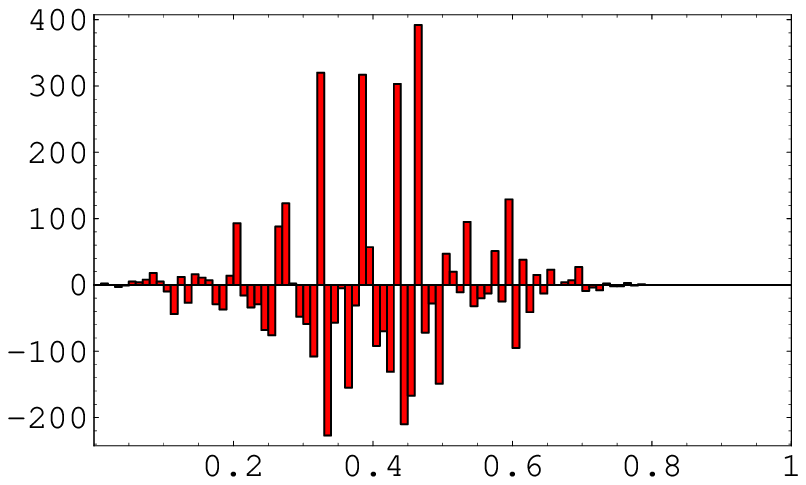}
\includegraphics[width=3.5cm]{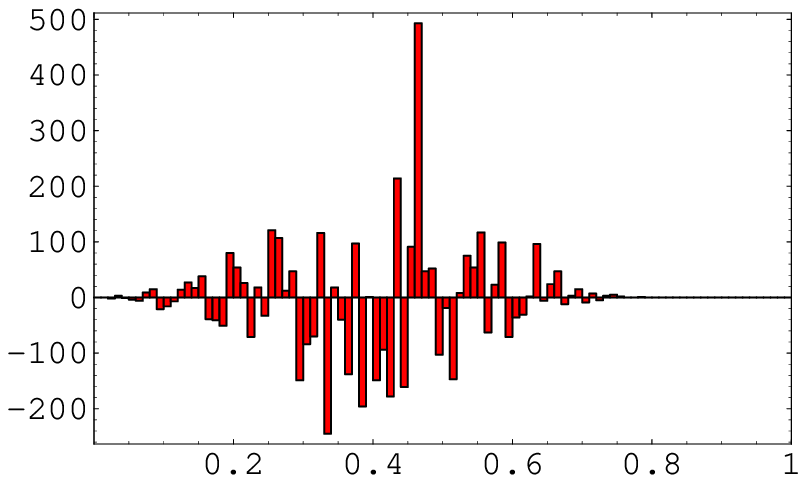}
\includegraphics[width=3.5cm]{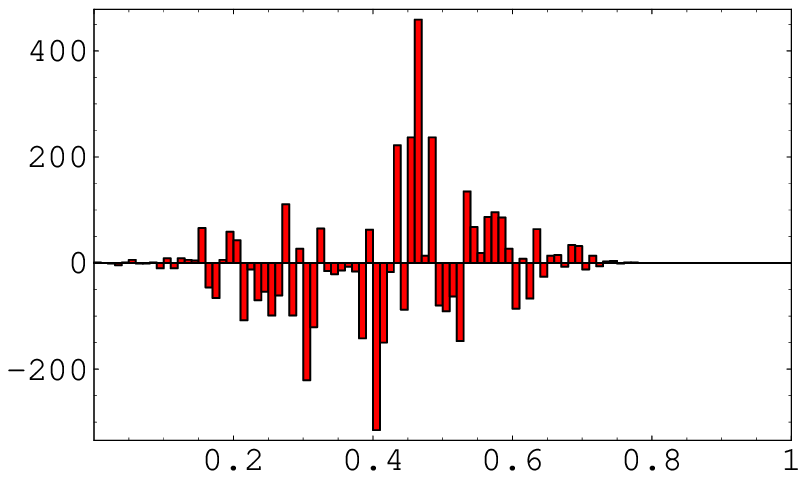}
\includegraphics[width=3.5cm]{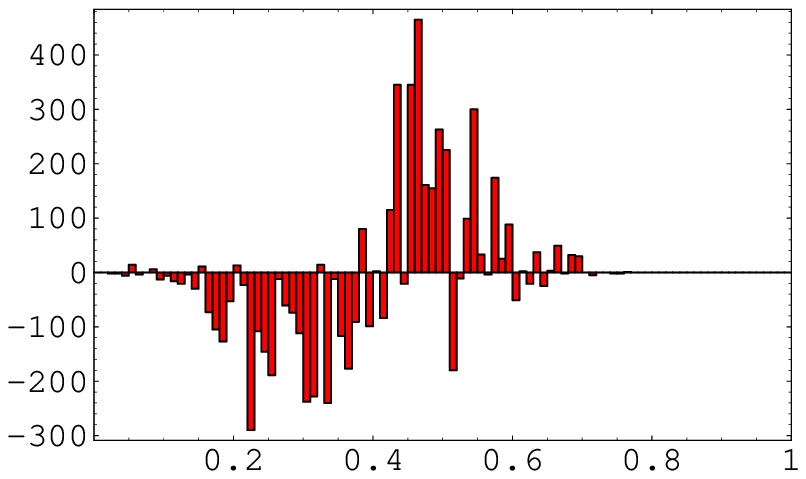}

\caption{Differences between the DSC contaminated distribution and 
the uncontaminated one for the WMAP amplitude for the alignment Hexadecapole-Quadrupole
in the case of small $\alpha$, vanishing $\beta$ and nominal scanning strategy. 
First row: $D42$. Second row: $S42$. From left to right (in every row) $p=1/1000$, $p=4/1000$, $p=7/1000$, $p=10/1000$. All the panels present the counts ($y$-axis) versus the statistic ($x$-axis).
See also the text.}

\label{fig10}
\end{figure*}

\begin{figure*}

\centering

\includegraphics[width=3.5cm]{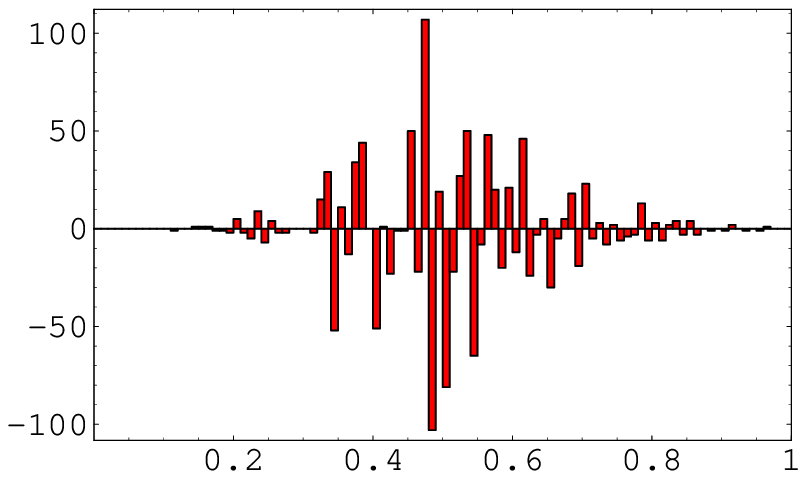}
\includegraphics[width=3.5cm]{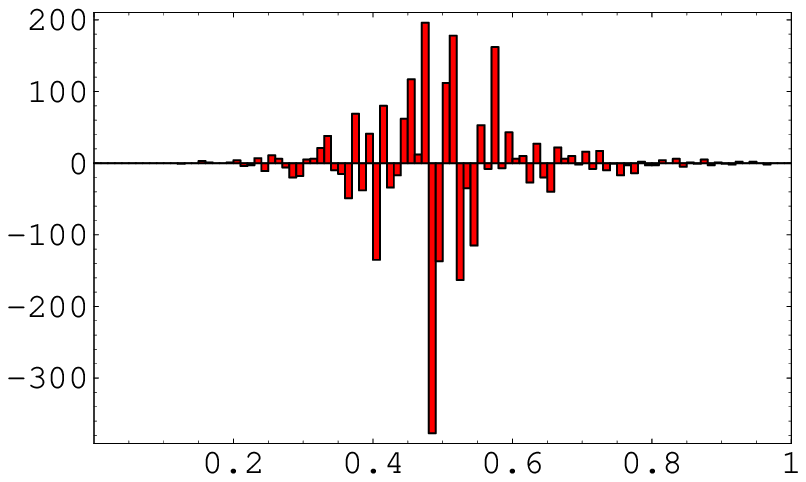}
\includegraphics[width=3.5cm]{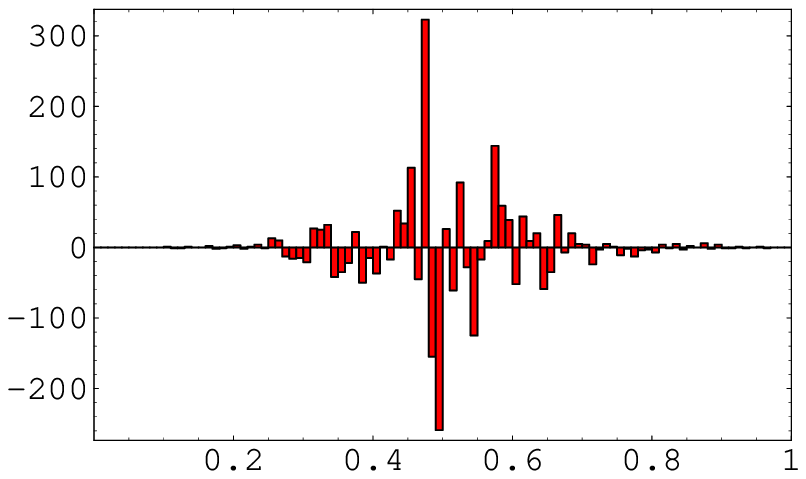}
\includegraphics[width=3.5cm]{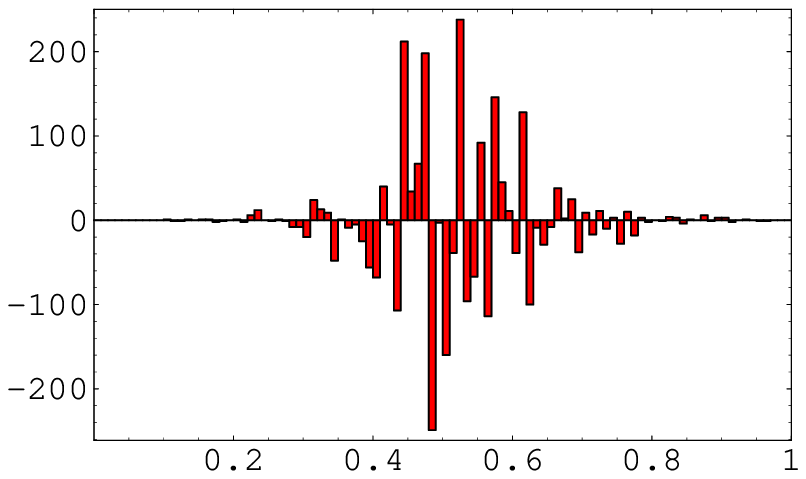}

\includegraphics[width=3.5cm]{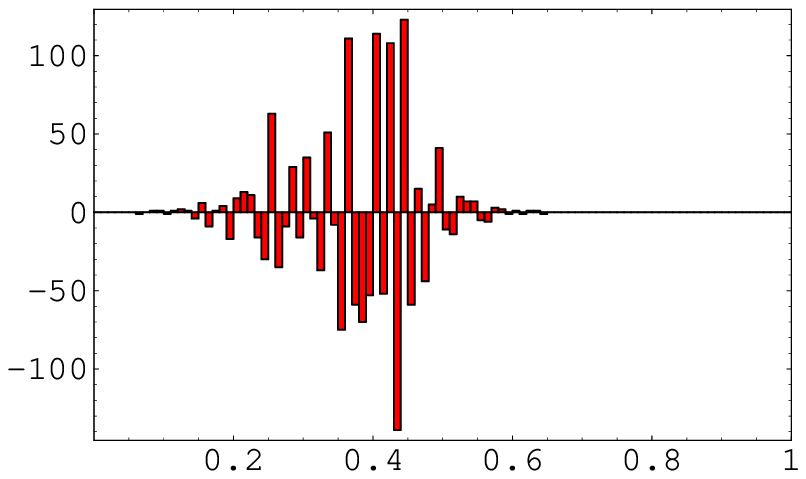}
\includegraphics[width=3.5cm]{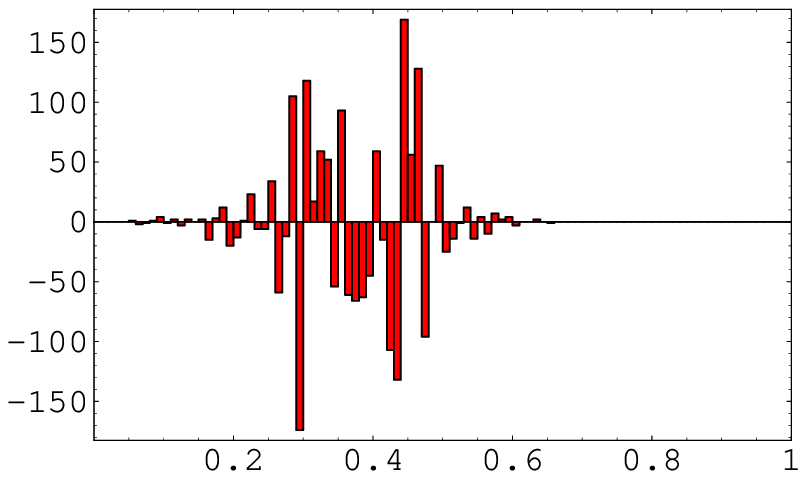}
\includegraphics[width=3.5cm]{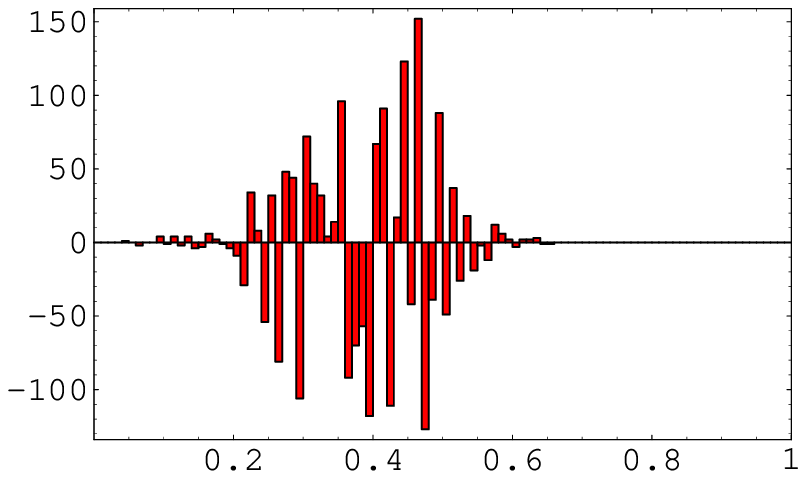}
\includegraphics[width=3.5cm]{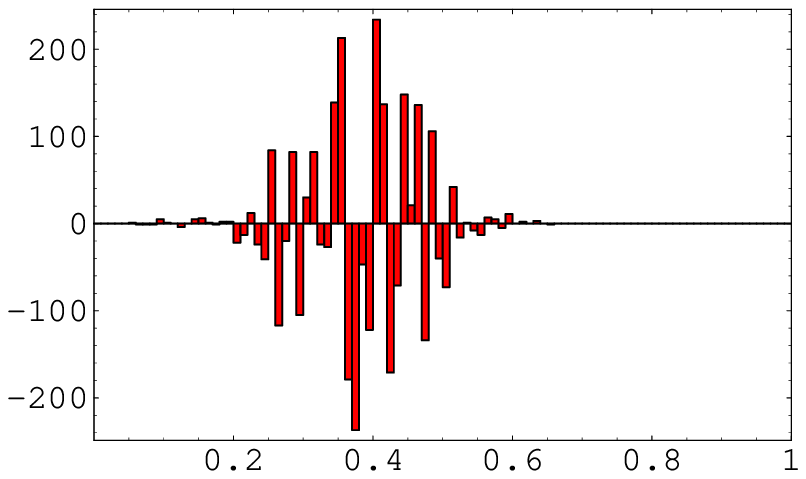}

\caption{Differences between the DSC contaminated distribution and 
the uncontaminated one in $\Lambda$CDM model for the alignment Hexadecapole-Octupole
in the case of small $\alpha$, vanishing $\beta$ and nominal scanning strategy. 
First row: $D43$. Second row: $S43$. From left to right (in every row) $p=1/1000$, $p=4/1000$, $p=7/1000$, $p=10/1000$.All the panels present the counts ($y$-axis) versus the statistic ($x$-axis).
See also the text.}

\label{fig11}
\end{figure*}

\begin{figure*}

\centering

\includegraphics[width=3.5cm]{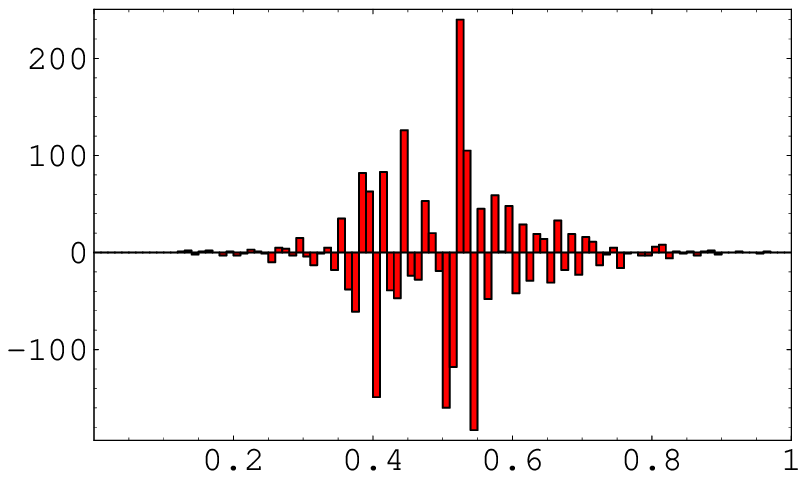}
\includegraphics[width=3.5cm]{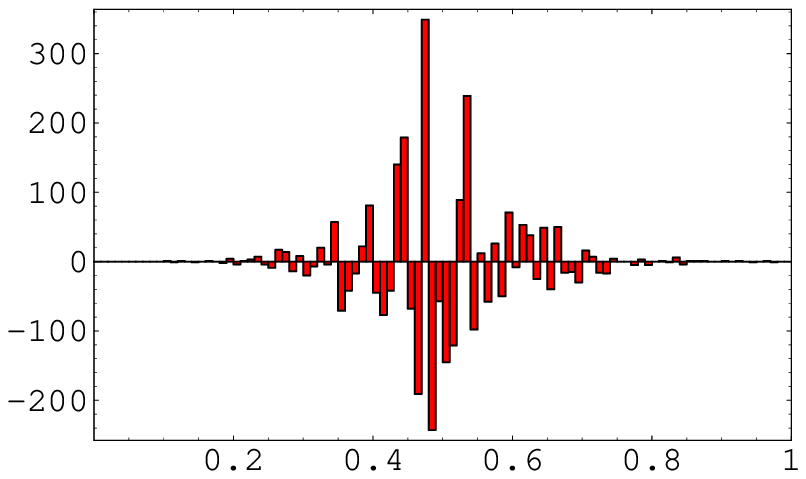}
\includegraphics[width=3.5cm]{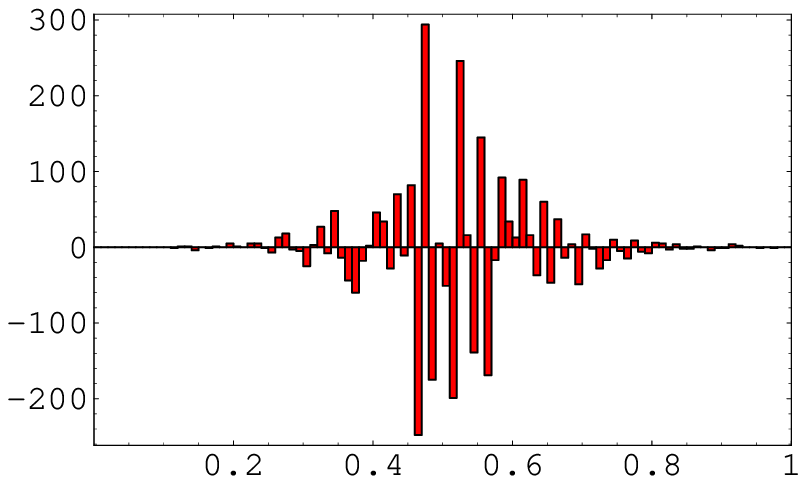}
\includegraphics[width=3.5cm]{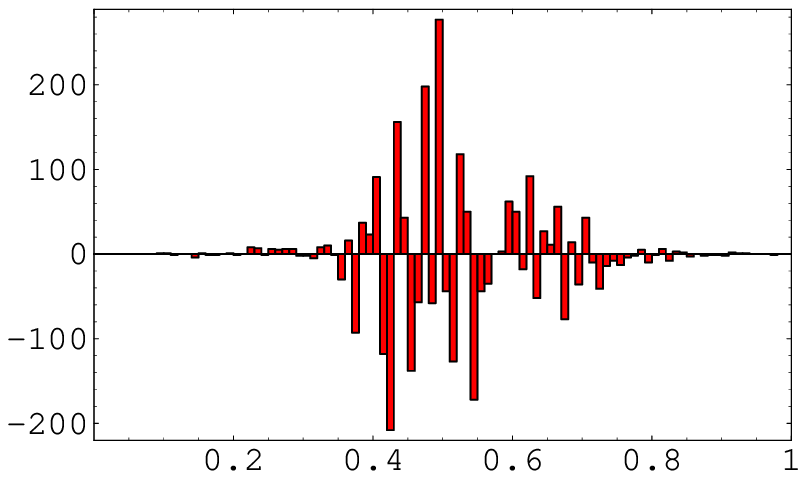}

\includegraphics[width=3.5cm]{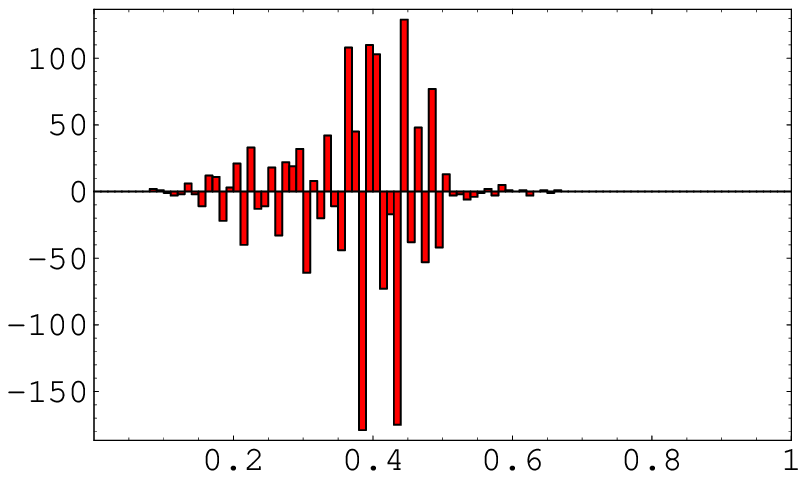}
\includegraphics[width=3.5cm]{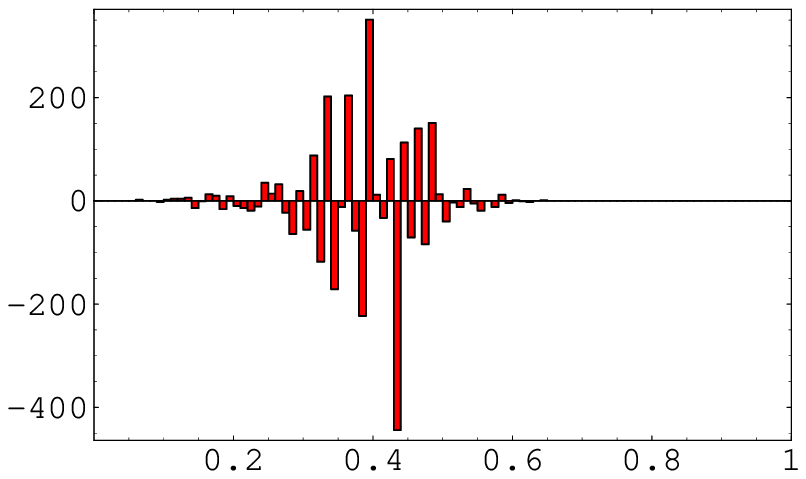}
\includegraphics[width=3.5cm]{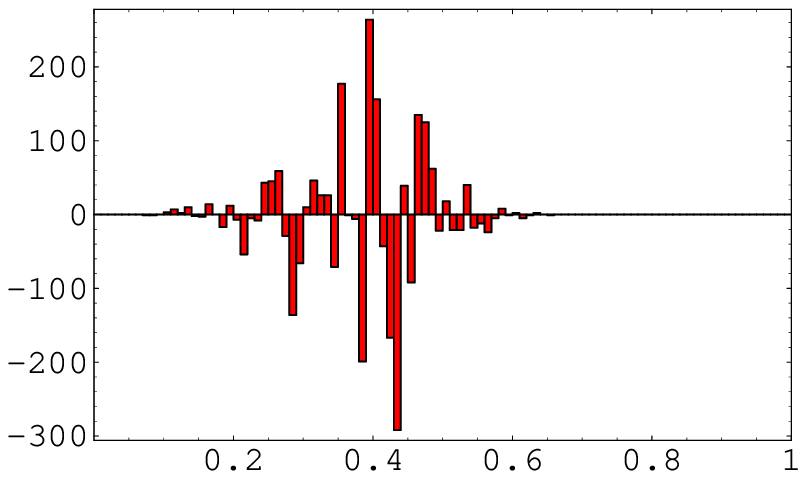}
\includegraphics[width=3.5cm]{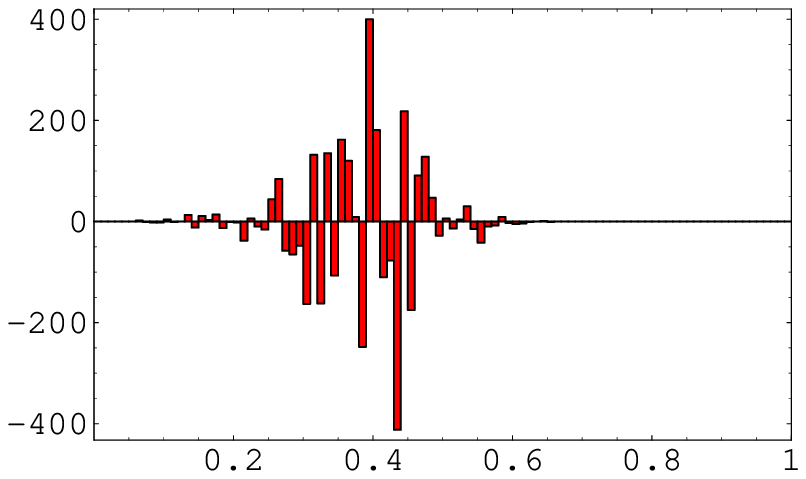}

\caption{Differences between the DSC contaminated distribution and 
the uncontaminated one for the WMAP amplitude for the alignment Hexadecapole-Octupole
in the case of small $\alpha$, vanishing $\beta$ and nominal scanning strategy. 
First row: $D43$. Second row: $S43$. 
From left to right (in every row) $p=1/1000$, $p=4/1000$, $p=7/1000$, $p=10/1000$.
All the panels present the counts ($y$-axis) versus the statistic ($x$-axis).
See also the text.}

\label{fig12}
\end{figure*}

In all the subsequent figures we display the difference (DD) between the distribution
contaminated by DSC and the uncontaminated one:
Figs \ref{fig2}--\ref{fig14} refer to the considered NSS.
The Figures for the considered CSS are not shown being very similar
to the NSS. However some (small) differences are reported in
Section \ref{conclusion}.
Of course, the sum of counts of the uncontaminated distribution is always
$3\times10^5$ while the sum of counts of the DD is zero.

The DD for $\hat W2$, $W2$, $W2^{\circ}$ are shown in Fig. \ref{fig2}
for the considered $\Lambda$CDM model and in Fig. \ref{fig3}
for the model with WMAP amplitude.
For $\hat W3$, $W3$, $D33$ and $S33$ the DD is exactly zero for the considered 
NSS. Therefore they are not reported for sake of conciseness. 
For the considered CSS they are noisy-like.
The DD for $R22$ are plotted in Fig.\ref{fig4} for the considered
$\Lambda$CDM model and for the WMAP amplitude.
The DD for $D23$ and $S23$ are shown in Fig. \ref{fig5} for the considered $\Lambda$CDM model
and in Fig. \ref{fig6} for the WMAP amplitude.
The DD for $\hat W4$ and $W4$ are plotted in Fig. \ref{fig7} for the
considered $\Lambda$CDM model and in Fig. \ref{fig8} for the WMAP amplitude.
The DD for $D42$ and $S42$ are given in Fig. \ref{fig9} for the considered $\Lambda$CDM model 
and in Fig. \ref{fig10} for the WMAP amplitude.
The DD for $D43$ and $S43$ are shown in Fig. \ref{fig11} for the considered $\Lambda$CDM model
and in Fig. \ref{fig12} for th WMAP amplitude.
The DD for $D44$ and $S44$ are plotted in Fig. \ref{fig13} for the considered $\Lambda$CDM model
and in Fig. \ref{fig14} for the WMAP amplitude.
The not reported DD for NSS are essentially noisy-like.



\section{Discussion and conclusions}
\label{conclusion}
\setcounter{equation}{0}

In this paper we have studied the impact of a non-proper subtraction of DSC on
alignments of multipole vectors associated to low multipoles for the forthcoming 
{\it Planck} mission.
This work represents a step forward of 
BGF06
in the study of DSC on the low multipoles of CMB pattern.

The analyses presented in this paper and 
in BGF06
provide a working 
example of a possible connection between the two low $\ell$ anomalies 
(APS and alignments of the low multipole vectors) 
in the presence of a non cosmological residual in the data.
To our knowledge, a non-properly subtracted systematic effect 
(of instrumental or astrophysical origin, or, as in this case,
coming from a combination of them)
represents the easiest way to link the statistics of $a_{\ell m}$ and the amplitude of $C_{\ell}$,
otherwise disconnected since the symmetry defined in equations (\ref{almandc}) and (\ref{Aandc})
holding at least for any distribution writable as in footnote 7 (like the Gaussian one).

We summarize the main results of this paper separately for 
the considered nominal and cycloidal scanning
strategies in the context of the {\it Planck} mission \citep{dupac}. 

\begin{figure*}

\centering

\includegraphics[width=3.5cm]{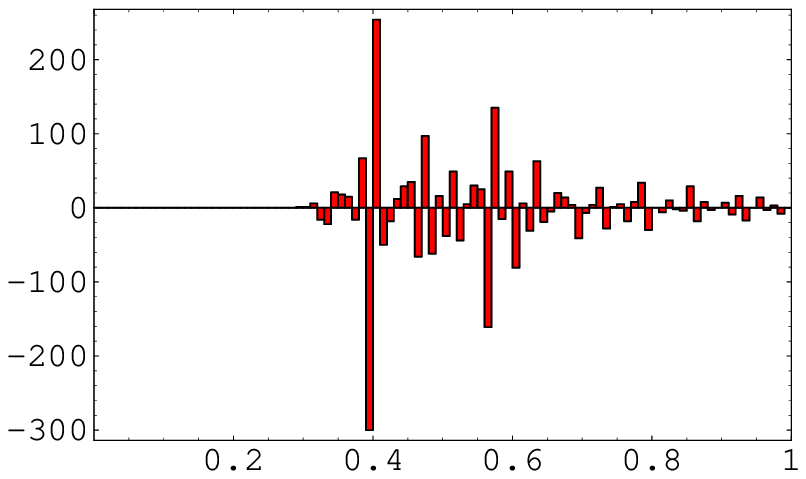}
\includegraphics[width=3.5cm]{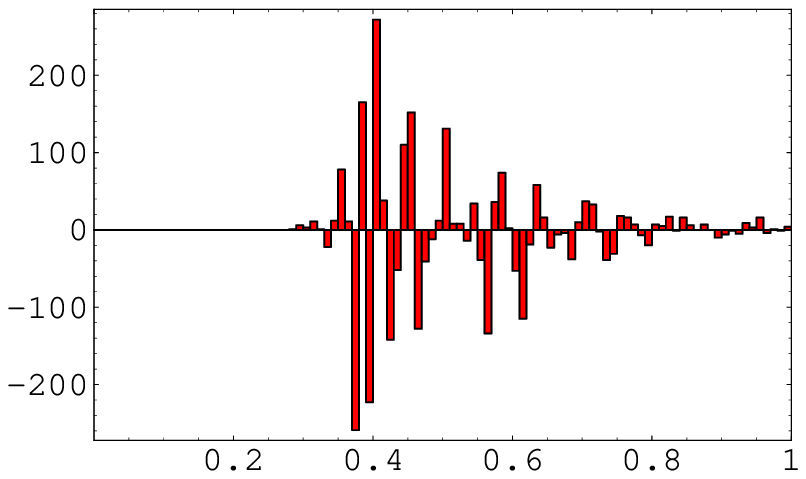}
\includegraphics[width=3.5cm]{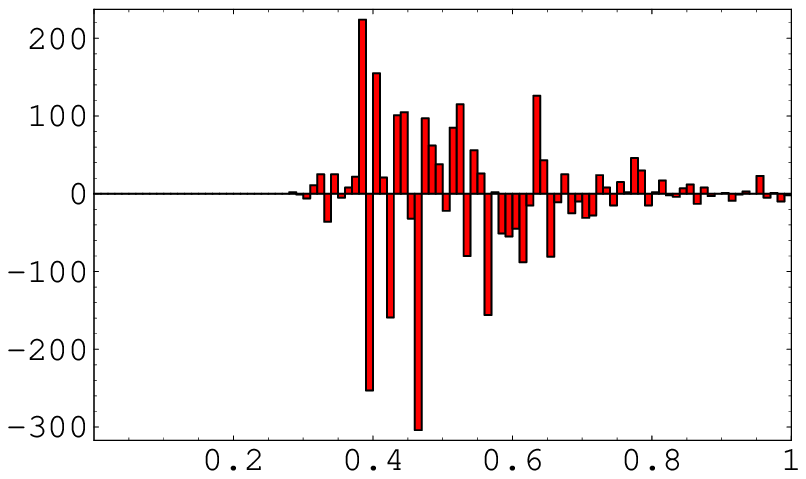}
\includegraphics[width=3.5cm]{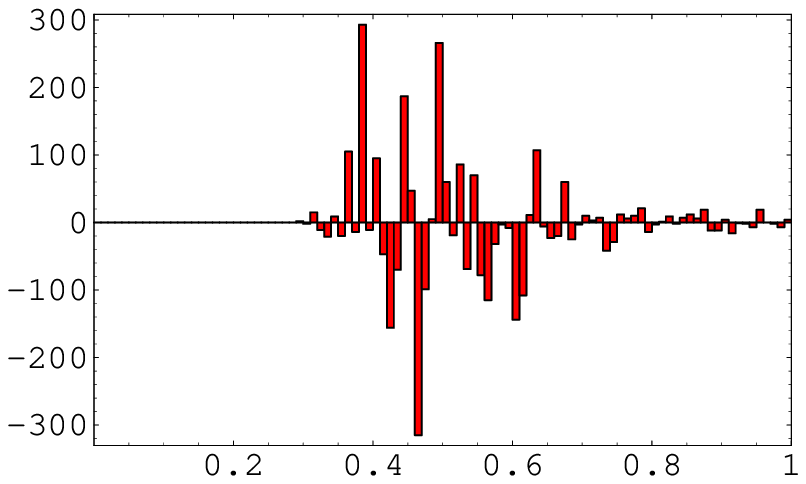}

\includegraphics[width=3.5cm]{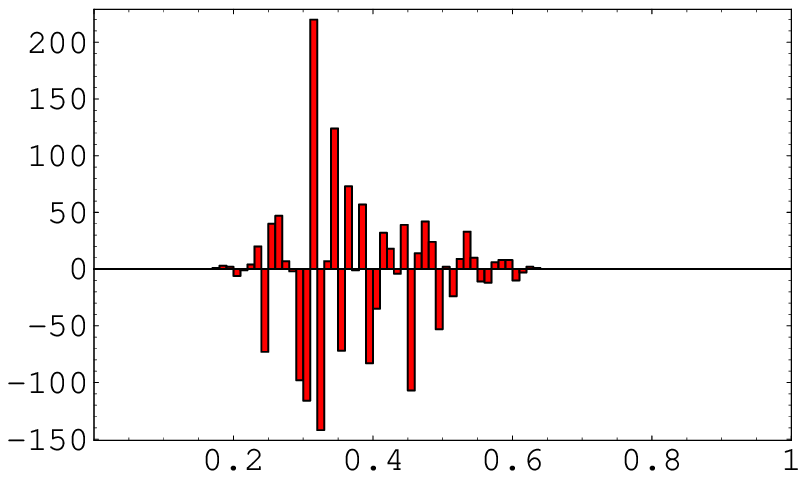}
\includegraphics[width=3.5cm]{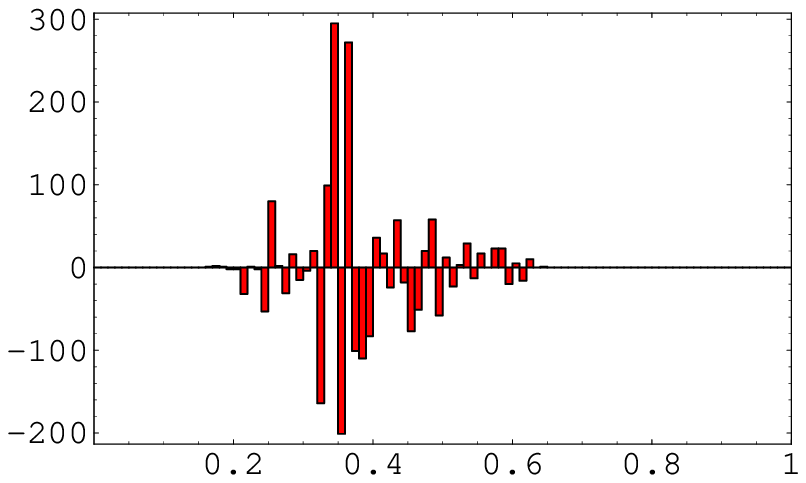}
\includegraphics[width=3.5cm]{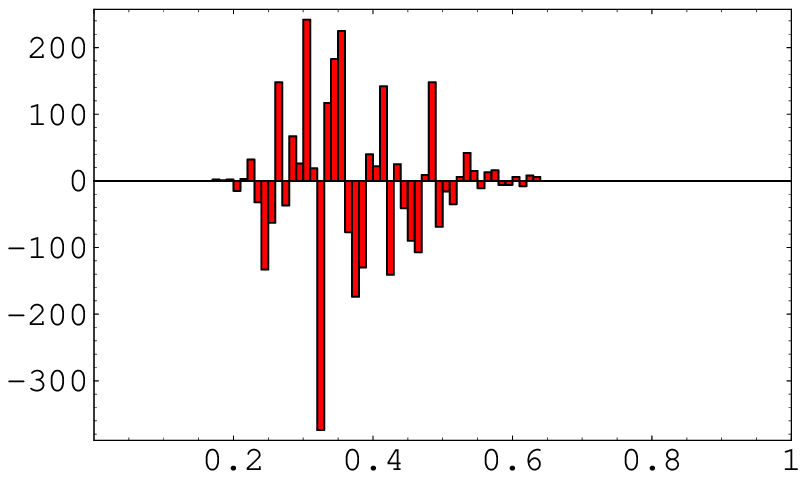}
\includegraphics[width=3.5cm]{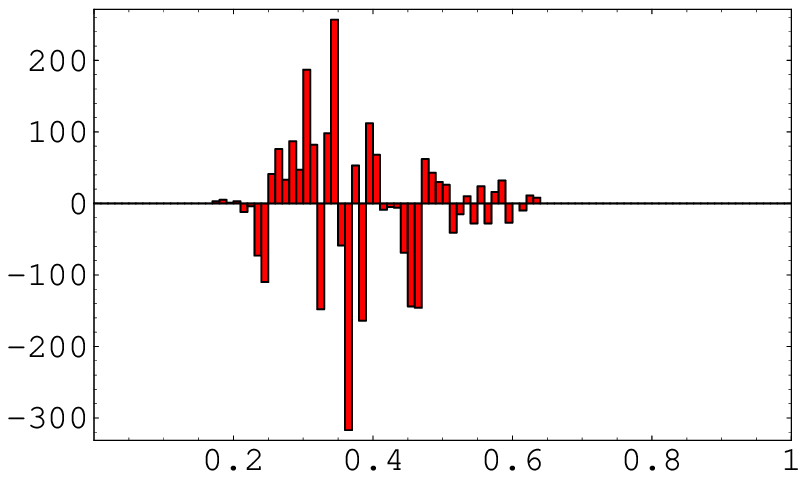}

\caption{Differences between the DSC contaminated distribution and 
the uncontaminated one in $\Lambda$CDM model for the self-alignment of the Hexadecapole
in the case of small $\alpha$, vanishing $\beta$ and nominal scanning strategy. 
First row: $D44$. Second row: $S44$. 
From left to right (in every row) $p=1/1000$, $p=4/1000$, $p=7/1000$, $p=10/1000$.
All the panels present the counts ($y$-axis) versus the statistic ($x$-axis).
See also the text.}

\label{fig13}

\end{figure*}

\begin{figure*}

\centering

\includegraphics[width=3.5cm]{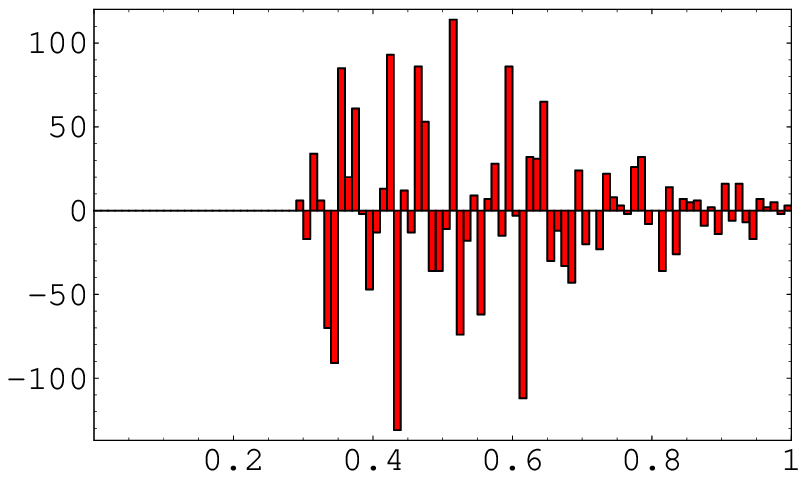}
\includegraphics[width=3.5cm]{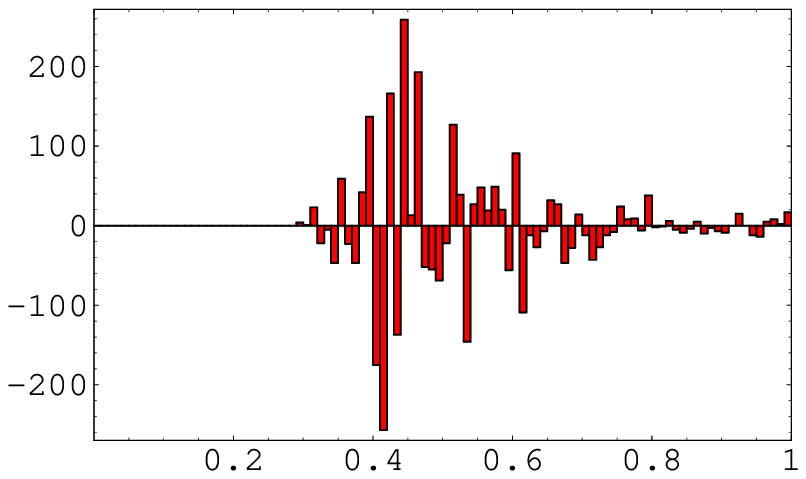}
\includegraphics[width=3.5cm]{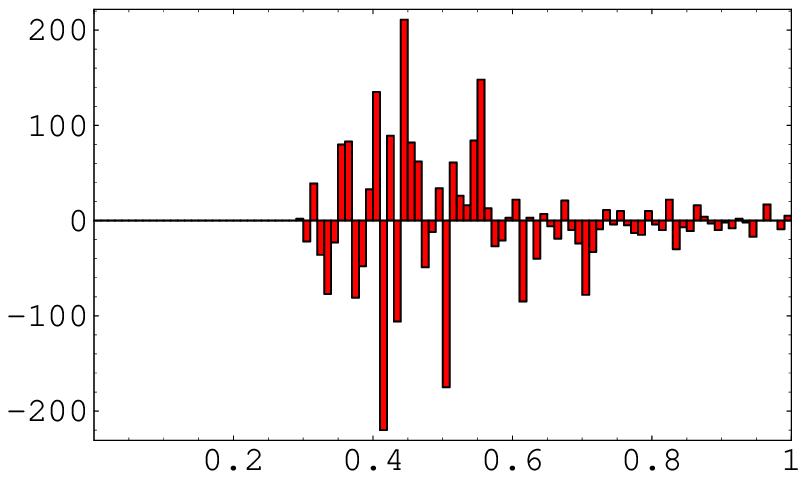}
\includegraphics[width=3.5cm]{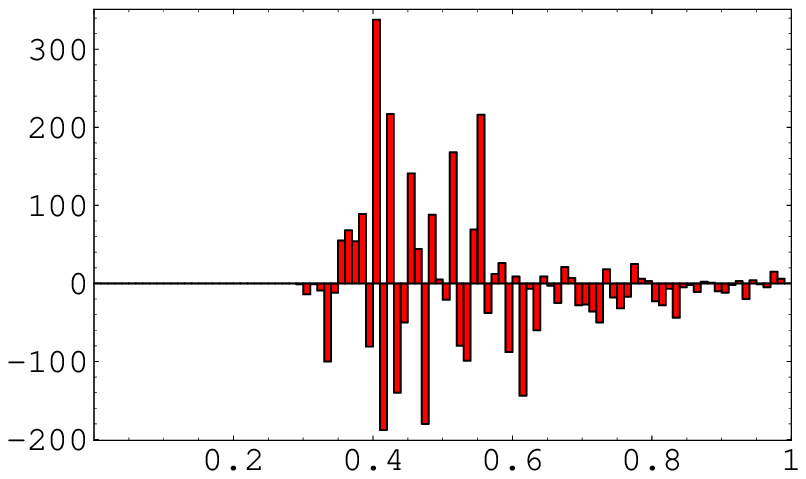}

\includegraphics[width=3.5cm]{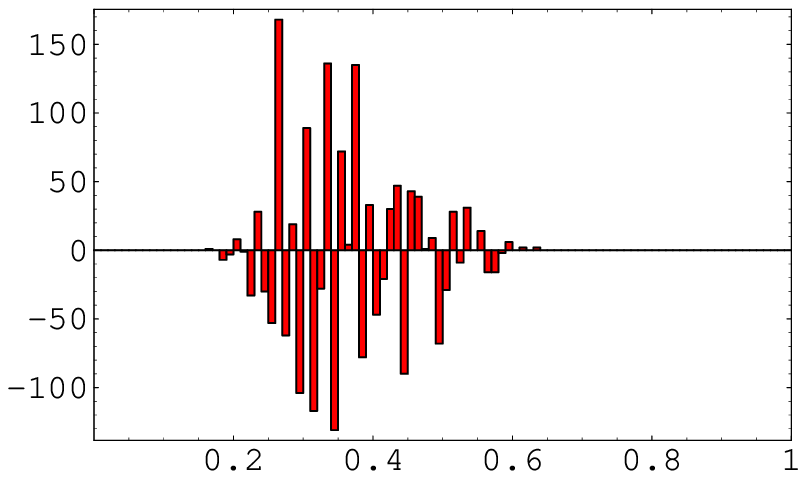}
\includegraphics[width=3.5cm]{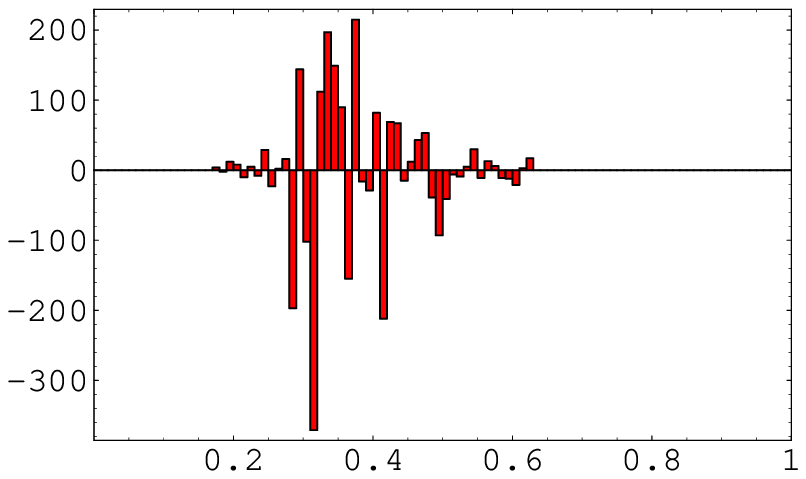}
\includegraphics[width=3.5cm]{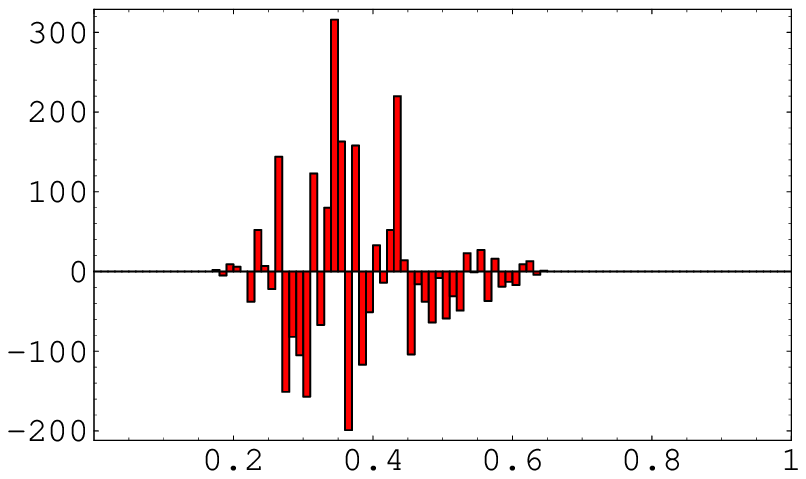}
\includegraphics[width=3.5cm]{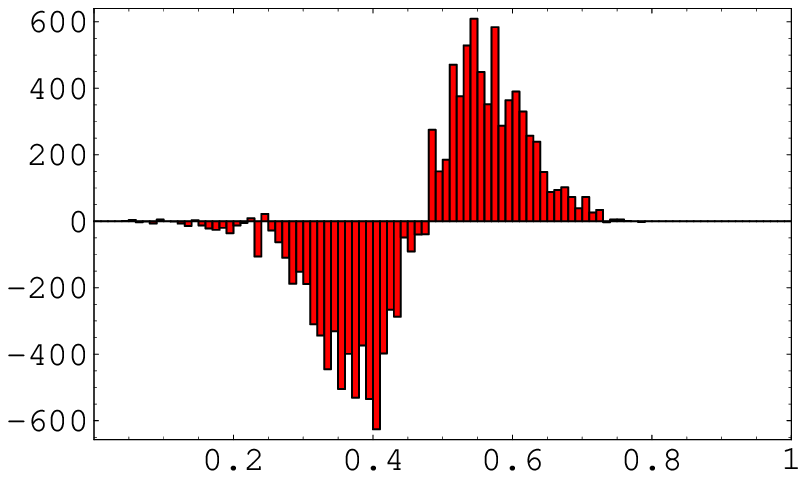}

\caption{Differences between the DSC contaminated distribution and 
the uncontaminated one for the WMAP amplitude for the self-alignment of the Hexadecapole
in the case of small $\alpha$, vanishing $\beta$ and nominal scanning strategy. 
First row: $D44$. Second row: $S44$. 
From left to right (in every row) $p=1/1000$, $p=4/1000$, $p=7/1000$, $p=10/1000$.
All the panels present the counts ($y$-axis) versus the statistic ($x$-axis).
See also the text.}

\label{fig14}

\end{figure*}

\smallskip
For the NSS:

\begin{itemize}
\item the probability of alignment quadrupole-dipole 
($\hat W2 $,$ W2 $,$ W2 ^{\circ}$)
tends to be lowered; 
the impact of this effect increases going from $\Lambda$CDM-like APS amplitudes 
to WMAP-like APS amplitudes and for increasing values of $p$, 
the percentage of power entering the main spillover with respect to the main beam 
(see Figs \ref{fig2} and \ref{fig3}); 
\item some features show up in the estimator for the alignment hexadecapole-dipole 
($\hat W4 $,$ W4 $) if $p$ is sufficiently large, {\it both} for $\Lambda$CDM 
and WMAP-like amplitudes (see Figs \ref{fig7} and \ref{fig8}); 
\item some features show up in the estimator for the alignment hexadecapole-quadrupole 
($D42$,$S42$) but only for
large $p$, i.e. $p=1/100$, {\it and} WMAP-like intrinsic amplitude
(see Fig. \ref{fig10});
\item the remaining estimators 
($R22$, $D23$, $S23$, $D42$, $S42$, $D43$, $S43$, $D44$, $S44$)
do not show remarkable features, being essentially noisy-like
(see Figs \ref{fig5}, \ref{fig9}, \ref{fig11}, \ref{fig12}, \ref{fig13} and \ref{fig14}); 
on the other hand, a weak signature appears in the case of $R22$, $S23$ and $S44$ (for the 
alignment octupole-quadrupole and self-alignment of the hexadecapole, respectively) 
for WMAP-like intrinsic amplitudes and the maximum considered value of $p$ (see Figs 
\ref{fig4}, \ref{fig6} and \ref{fig14}). 

\end{itemize}

For the CSS:
\begin{itemize}
\item the alignment quadrupole-dipole 
($\hat W2 $,$ W2 $,$ W2 ^{\circ}$)
again tends to be lowered, 
the impact of this effect being again stronger if the intrinsic sky 
amplitude is lower and $p$ is sufficiently large
;
\item some features appear for the estimator of the alignment hexadecapole-dipole 
($\hat W4 $,$ W4 $)
but stronger than those obtained for the NSS and increasing with
$p$ 
; 
\item there is a clear signature in the estimator of the alignment 
hexadecapole-quadrupole 
($D42$,$S42$)
but only for WMAP-like amplitudes 
;
\item again, the remaining estimators 
($R22$, $D23$, $S23$, $D43$, $S43$, $D44$, $S44$)
do not show remarkable features, being essentially noisy-like;
on the other hand, a weak signature appears also for this scanning strategy
in the case of $R22$, $S23$ and $S44$ (self alignment of the quadrupole, alignment octupole-quadrupole and self alignment of the hexadecapole respectively),
for WMAP-like intrinsic amplitudes
and the maximum considered value of $p$.

\end{itemize}

We conclude that possible residual DSC should leave a non-negligible
impact on low multipole alignments 
for far sidelobe integrated responses 
corresponding to effective values of $p$ 
$\gsim {\rm few} \times 10^{-3}$.

We note also that, in general, it could be very useful to 
carry out the alignment analysis by exploiting 
both normalized and unnormalized estimators, since they
could show different sensitivity in the diagnostic of 
the same effect, as evident for example 
in Fig. \ref{fig5}.

The scanning strategy and the behaviour of the optics of WMAP
are significantly different from those considered here, suitable 
for {\it Planck}.
It is then difficult to extrapolate the above results to the
WMAP surveys. 
If the large scale footprint of the DSC
is not so critically dependent on the above details
and our analysis applies as well to 
the WMAP 
surveys,
DSC may not easily explain  
the whole set of
anomalous alignments at large scale 
found also in 
WMAP three year release (see \cite{copi3y}). 
However, we believe that further studies of this and
other systematic effects are required.

Provided that 
the real sidelobes of the {\it Planck} receivers in flight conditions 
will correspond to values of $p$ 
$\lsim {\rm few} \times 10^{-3}$
-- as realistically expected \citep{sandri} 
at least in the cosmological frequency channels --
and will be known with relative accuracies
better than $\sim {\rm few} \times 10$\%
(leaving to smaller residual contaminations,
equivalent to $p \lsim 10^{-3}$,
after a suitable cleaning during
data reduction)
{\it Planck} maps will be very weakly affected 
by DSC on the alignments. 

\bigskip

\newpage

\noindent
{\bf Acknowledgements.}
We gratefully thank J.~Weeks for useful discussions.
We warmly acknowledge all the members of the {\sc Planck}
Systematic Effect Working Group for many conversations and collaborations.
It is a pleasure to thank D.~Maino and P.~Naselsky
for stimulating conversations.
We ackowledge the use of the codes for the computation of multipole vectors
provided by \cite{weeks}.
Some of the results in this paper have been derived using {\tt HEALPix}
\citep{gorski05}. 
The use of the WMAP three year release data products is acknowledged.
 {\it This work has been done in the framework of the
{\it Planck} LFI activities.}


\end{document}